\documentstyle[aps,pre]{revtex}
\begin{document}
\draft
\preprint{SNUTP 95-112}
\title{Finite-size scaling and the toroidal partition function \\
of the critical asymmetric six-vertex model}
\author{Jae Dong Noh and Doochul Kim }
\address{Department of Physics and Center for Theoretical Physics,
 Seoul National University,\\ Seoul 151-742, Korea}
\maketitle
\date{\today}
\begin{abstract}
Finite-size corrections to the energy levels of the asymmetric six-vertex
model transfer matrix are considered using the Bethe ansatz solution for
the critical region.
The non-universal complex anisotropy factor is related to the bulk
susceptibilities. The universal Gaussian coupling constant $g$ is also
related to the bulk susceptibilities as $g=2H^{-1/2}/\pi$,
$H$ being the Hessian of the bulk free energy surface viewed
as a function of the two fields.
The modular covariant toroidal partition function is derived in
the form of the  modified Coulombic partition function which embodies the
effect of incommensurability through two mismatch
parameters. The effect of twisted boundary conditions is also considered.
\end{abstract}

\section{Introduction}\label{sec1}
The six-vertex model was first introduced as a model for the residual
entropy of ice and for related ferroelectric transitions~\cite{LieW}
but more recently, several other physical applications are being found.
In the body-centered solid-on-solid models, the six-vertex configuration
is mapped to the surface configuration of the fcc and bcc faces and the
free energy of the six-vertex model as a function of the horizontal and
vertical fields depicts the equilibrium crystal shapes~(ECS)~\cite{ecs,BukS}.
Also, on the stochastic surface where the vertex weights satisfy a
certain relation, the transfer matrix of the six-vertex model can be
regarded as the transition matrix of probabilistic cellular automata
describing the dynamics of a driven lattice gas system and a (1+1) dimensional
surface growth model~\cite{GwaS,NeeN,Kim}. The five-vertex model, which is a
special asymmetric limit of the six-vertex model, can be viewed as models
for an interacting domain wall system~\cite{NK1}, interacting
dimers~\cite{Hwang}, and certain types of crystal
surfaces~\cite{GulBL}. In these applications, it is necessary to consider the
asymmetric six-vertex~(ASSV) model in which there are non-zero horizontal and
vertical fields which break the arrow reversal symmetries. It becomes
the symmetric six-vertex~(SSV) model when the fields are zero.
The Heisenberg XXZ chain is a closely related problem.
The Hamiltonian for the XXZ chain can be
obtained from an anisotropic limit of the transfer matrix of the six-vertex
model~\cite{Kim}. The vertical field corresponds to the magnetic field which
couples to the $z$ component of spins and the horizontal field corresponds to
the asymmetry in the two hopping rates. The XXZ chain is called symmetric
(asymmetric) if the two hopping rates are same (different).

The phase diagram and the nature of the phase transition in the ASSV model
are well known~\cite{LieW,Nol}.
The phase diagram consists of the ferroelectrically and
anti-ferroelectrically ordered phases and the disordered phase.
Especially, the disordered phase shows interesting scaling properties.
It is the critical phase with continuously varying critical
exponents and is described by the
central charge $c$=1 conformal field theory (CFT) in the continuum
limit~\cite{Car}. Since the $c$=1 theory plays a basic role in the
theory of two dimensional critical phenomena,
there have been many works on the scaling properties of the six-vertex
model. However, most of these works are confined to the SSV
model~\cite{FraSZ_Pasquer} or the symmetric XXZ chain with or without the
magnetic field~\cite{Alcaraz,BogIK_BogIR,WoyET,ParkW} and to the special
case of the five-vertex model~\cite{NK1}, while the critical properties of the
general ASSV model are also of interest.  One purpose of this work is to fill
this gap. In this paper, we investigate the critical properties of the ASSV
model through the finite-size scaling~(FSS) studies on the transfer matrix
spectra. Our work also covers the critical properties of the asymmetric XXZ
chain.

The FSS is very useful in studying the critical behaviors of two-dimensional
statistical systems on lattice~\cite{PrivCardy}.
When the system is in a critical phase which possesses the conformal symmetry,
it imposes strong restrictions on the FSS form of the eigenvalue spectra of
the transfer matrix. Let ${\bf T}$ denote the row-to-row transfer matrix for
a system with $N$ columns and $M$ rows.
We assume periodic boundary conditions unless stated otherwise.
The energy of the level $\alpha$ defined by
$E_\alpha = - \ln \Lambda_\alpha$, $\Lambda_\alpha$ being the
$\alpha$-th eigenvalue of ${\bf T}$, is expected to follow the scaling form
\begin{equation}\label{e.fss.crt}
{E_\alpha} = Nf  +\frac{2\pi}{N} \zeta'' \left(h_\alpha+
\overline{h}_\alpha-\frac{c}{12} \right) - \frac{2\pi i}{N} \zeta'
\left(h_\alpha-\overline{h}_\alpha\right) + o\left(\frac{1}{N}\right)
\end{equation}
where $f$ is the bulk free energy (in units of $k_B T$), $\zeta'$ ($\zeta''$)
is the real (imaginary) part of the complex anisotropy factor $\zeta=
\zeta' + i \zeta ''$,
$c$ is the central charge, $(h_\alpha,\overline{h}_\alpha)$ are
the conformal dimensions associated with the level and $o(x)$ stands for
terms smaller than $x$; $\lim_{x\rightarrow 0} o(x)/x =0$.
The level whose energy follows the scaling form of Eq.~(\ref{e.fss.crt}) is
associated with a primary operator and its descendants of the corresponding
CFT.  When the one lattice-unit translation by ${\bf T}$ is incommensurate
with the periodicity of the underlying model, an imaginary term of $O(1)$
may appear in the right-hand side of Eq.~(\ref{e.fss.crt})~\cite{KimP}.
The toroidal partition function (TPF) $\widetilde{\cal Z}$ is defined as the
 $O(1)$ part of the partition function:
\begin{equation}
\widetilde{\cal Z} = \lim_{\stackrel{N,M\rightarrow\infty}{M/N={\text{fixed}}
  }}
\sum_\alpha e^{-M(E_\alpha-Nf)}
\end{equation}
where the sum is over all levels.
 If we use the scaling form of Eq.~(\ref{e.fss.crt}),
it takes the form
\begin{equation}
\widetilde{\cal Z} = {({\sf q}\bar{\sf q})^{-\frac{c}{24}}}
\sum_{\alpha} {\sf q}^{h_\alpha} \bar{\sf q}^{\overline{h}_\alpha}
\end{equation}
where the
nome ${\sf q}\equiv e^{2\pi i\tau}$ with  $\tau =\tau'+i\tau''=
 \frac{M}{N}\zeta$ and $\bar{\sf q}$ is the complex conjugate of
${\sf q}$. The complex parameter
$\tau$ is the modular ratio of the torus on which the corresponding
CFT is defined and  specifies how
the $N\times M$ lattice should be deformed to make the system
isotropic in the continuum limit.
The TPF contains the whole information on
the spectra or the operator content of the model
  and enjoys the modular invariance
(or more generally, modular covariance) properties which together with the
conformal invariance principle is often sufficient to determine the form of
 $\widetilde{\cal Z}$ ~\cite{Car86_ZuberCap}.
The $c$=1 CFT consists of
three isolated points and two one-parameter families~\cite{Ginsparg}.
The latter two
describe the critical eight-vertex model or the Ashkin--Teller model and
the SSV model or the Gaussian model compactified on a circle,
respectively. They are related by a duality
transformation.  The TPF of the $c$=1 theory corresponding to the
SSV line is the so-called Coulombic partition function~\cite{FraSZ_Pasquer}
given by, for $M$ and $N$ even,
\begin{equation} \label{coulomb}
\widetilde{\cal Z}_C = \frac{1}{|\eta({\sf q})|^2} \sum_{m,n\in {\bf Z}}
   {\sf q}^{\Delta_{m,n}} \bar{{\sf q}}^{\overline{\Delta}_{m,n}}
\end{equation}
where $\eta({\sf q})$ is the Dedekind eta function,
\begin{equation} \label{dedekind}
 \eta ({\sf q}) = {\sf q}^{1/24} \prod_{n=1}^\infty (1-{\sf q}^n)
\end{equation}
and the conformal dimensions are given by
\begin{equation}\label{e.opc.g}
\begin{array}{ccl}
\Delta_{m,n} &=& \frac{1}{4}\left(\frac{m}{\sqrt{g}}+\sqrt{g}n\right)^2 \\
\overline{\Delta}_{m,n}
 &=& \frac{1}{4}\left(\frac{m}{\sqrt{g}}-\sqrt{g}n\right)^2 .
\end{array}
\end{equation}
Here $g$ is the Gaussian coupling constant and the integer indices $m$ and $n$
label the spin-wave
excitation and  the vortex excitation, respectively. The Gaussian coupling
constant $g$ is defined in such a way that it takes the value 1/2 for the
free fermion theory and is related to $K_R$ of Ref.~\cite{JoseKKN} by
$g=2\pi K_R$.
However, the CFT being a bootstrap theory, it does not tell us how $g$ and
$\tau$ are related to the lattice model parameters. To obtain that
information, one needs to rely on the FSS analysis.

The transfer matrix for the six-vertex model or equivalently the Hamiltonian
of the XXZ chain is diagonalized~\cite{LieW} by the Bethe ansatz method for
general boundary conditions.  In this paper, we present a method of
calculating the finite-size corrections of arbitrary low-lying energies
in the whole parameter space of the ASSV model.
Starting from the Bethe ansatz solution for the system of
width $N$, we derive a systematic expansion in $1/N$ of the energies
assuming certain analyticity properties of the phase function which is
introduced in Sec.~\ref{sec2}.
The method is very similar to that used in \cite{WoyET} which considered
the symmetric XXZ chain in a magnetic field but is generalized to be
applicable to the general cases.
{}From the expansion, we find that the whole critical phase of the ASSV model
is also in the Gaussian-model universality class with $c=1$, where the
coupling constant $g$ is obtained from the solution of an integral equation.
And we also find that though the susceptibilities are non-universal, i.e.,
they depend on non-universal parameters, the Hessian of the free
energy is simply given by
\begin{equation} \label{hessian}
  H \equiv \left|
  \begin{array}{cc}
    \frac{\partial^2 f}{\partial^2 h} &   \frac{\partial^2 f}{\partial h
 \partial v} \\   \frac{\partial^2 f}{\partial v \partial h} &
   \frac{\partial^2 f}{\partial^2 v} \end{array} \right|
 = \left(\frac{2}{\pi g}\right)^{2}
\end{equation}
and hence depends on the model parameters only through
$g$. Here, $h$ and $v$ are the horizontal and vertical fields, respectively,
in units of $k_B T$. Since we find the FSS form for a general class of energy
levels, we are able to construct the TPF for the ASSV model with the
periodic boundary conditions explicitly. The resulting expression is given by
the modified Coulombic partition function
\begin{equation}
\label{Z_mpf.a6v}
\widetilde{\cal Z} = \frac{1}{|\eta({\sf q})|^2} \sum_{m,n\in {\bf Z}}
   e^{-2 \pi i m \alpha}
   {\sf q}^{\Delta_{m,n-\beta}} \bar{{\sf q}}^{\overline{\Delta}_{m,n-
\beta}}
\end{equation}
where $\alpha$ and $\beta$ as defined in Eq.~(\ref{mismatch})
are the two mismatch parameters which account for
the incommensurability of the lattice with the mean distances between down
arrows and left arrows, respectively.
The ASSV model under the twisted boundary
conditions~(explained in Sec.~\ref{sec2}) is equivalent to the ASSV model
under the periodic boundary conditions with modified fields.
Using this relation the TPF for the ASSV model under the twisted boundary
conditions is also derived, which confirms the conjectured TPF for the SSV
model under the twisted boundary condition in the horizontal
direction~\cite{KimP2}.

This paper is organized as follows. In Sec.~\ref{sec2}, we give a brief review
of the transfer-matrix formulation and its Bethe ansatz solution of the ASSV
model. The classification scheme of the Bethe ansatz solutions is presented.
The relation between the ASSV model under the periodic and the twisted
boundary conditions is also discussed.
In Sec.~\ref{sec3}, under a certain assumption we derive a summation formula
which converts a sum over functions of rapidities for general levels into a
finite-size expansion.
The assumption turns out to be the sufficient condition for
the criticality in Sec.~\ref{sec4}.
We obtain  an integral equation for the expansion
coefficients and find that the finite-size correction terms are related to the
partial derivatives of the bulk term with respect to the horizontal and
vertical fields. Some of the details of calculations are relegated to
App.~\ref{app0}.
In Sec.~\ref{sec4}, applying the summation formula to the transfer matrix
eigenvalues, we derive the connections to the $c$=1 CFT together with the
expressions for $g$.  Since we find the FSS form for the general scaling
levels, the TPF's for the periodic and the general twisted boundary
conditions are derived.  In Sec.~\ref{sec5}, we summarize our
results and discuss relations to other works. Also, possible physical
relevances to the ECS are discussed.
In App.~\ref{app2} we prove an identity $J=1$ where $J$ appears in the
relations between the finite-size correction terms and susceptibilities.
In App.~\ref{app3} the modular transformation properties of the TPF for the
ASSV model with general boundary conditions are discussed.

\section{The asymmetric six-vertex model}\label{sec2}
On the square lattice of $N$ columns and of $M$ rows, the six-vertex model
configurations are obtained by covering the bonds of the square lattice with
arrows which satisfy the ice rule: At each vertex there
are two arrows in and two arrows out. The six vertex configurations
satisfying the ice rule are shown in Fig.~1.
Following the notation of Ref.~\cite{Nol}, the vertex energies are
assigned as
\begin{equation}\label{energy_assign}
\begin{array}{cclcccl}
\varepsilon_1 &=& -\frac{\delta}{2}-h-v,& &
\varepsilon_2 &=& -\frac{\delta}{2}+h+v \\
\varepsilon_3 &=& \frac{\delta}{2}-h+v, & &
\varepsilon_4 &=& \frac{\delta}{2}+h-v \\
\varepsilon_5 &=& -\epsilon, & &
\varepsilon_6 &=& -\epsilon \ .
\end{array}
\end{equation}
We write the energy in units of $k_B T$ and denote the vertex weights as
 $w_i=\exp(-\varepsilon_i)$. $h\ (v)$ is the horizontal
(vertical) electric field which is conjugate to the horizontal~(vertical)
polarization $1-2\rho_L$ ($1-2\rho_D$) where $\rho_L$ ($\rho_D$)
is the mean density of left~(down) arrows.
For later uses, we define  parameters
\begin{equation} \label{Delta}
\Delta= \frac{w_1 w_2 +w_3 w_4 - w_5 w_6}{2 \sqrt{w_1 w_2 w_3 w_4}}
 = \frac{
e^{\delta}+e^{-\delta}-e^{2\epsilon}}{2}
\end{equation}
and
\begin{equation} \label{delta_tilde}
\widetilde{\Delta} = \frac{w_1w_2+w_3w_4-w_5w_6}{w_1w_3+w_2w_4}
= \frac{\Delta}{\cosh (2h)} \ .
\end{equation}

The partition function ${\cal Z}$ is written as
\begin{equation}\label{e.Z.def}
{\cal Z} = {\rm Tr}\ {\bf T}^M \ ,
\end{equation}
where ${\bf T}$ is the row-to-row transfer matrix whose definition can be
found in Ref.~\cite{Nol}.
Let $Q$ denote the number of down arrows per row and let
 \begin{equation} \label{smallq}
  q= \frac{Q}{N}.  \end{equation}
The thermal average of $q$ is $\rho_D$.
Since $Q$ must be the same in the two adjacent
rows of vertical bonds due to the ice rule, ${\bf T}$
can be considered in each subspace of fixed number of $Q$, which will be
called the $Q$ sector, separately.
{}From the Bethe ansatz method~\cite{LieW}, the eigenvalue of ${\bf T}$ in
the $Q$ sector is given by
\begin{mathletters}
\begin{equation}
\Lambda = \Lambda_R+\Lambda_L
\end{equation}
where
\begin{eqnarray}
\Lambda_R&=&w_1^{N-Q}\prod_{j=1}^Q
\left(w_3-\frac{w_5w_6}{w_4-w_1z_j}\right)\ ,\\
\Lambda_L&=&w_4^{N-Q} \prod_{j=1}^Q\left(w_2+\frac{w_5w_6
         z_j}{w_4-w_1z_j}\right)
\end{eqnarray}
\end{mathletters}
where the fugacities $\{z_j\}$ are given by the solutions of the Bethe ansatz
equation
\begin{equation}\label{e.bae.bare}
z_i^N = (-1)^{Q-1} \prod_{j=1}^{Q} \frac{1 +e^{4h}z_iz_j-2e^{2h} \Delta z_i}{
1+e^{4h}z_iz_j-2e^{2h} \Delta z_j}\ .
\end{equation}
Note that the five-vertex model is achieved in the limit
$h\rightarrow \pm \infty$ keeping
$\widetilde{\Delta}$ fixed.
Introducing a variable $p$ such that $e^{ip}= z e^{2h}$ and the phase function
\begin{equation}\label{e.Z0.def}
Z_N^0(p) = p +2hi-\frac{1}{N} \sum_{j=1}^Q \Theta^0(p,p_j)
\end{equation}
where \begin{equation}\label{e.theta.def}
e^{i\Theta^0(p,p')}\equiv \frac{1 +e^{i(p+p')}-2\Delta e^{ip}}{
1+e^{i(p+p')}-2\Delta e^{ip'}}\ ,
\end{equation}
the Bethe ansatz equation takes the simple form
\begin{equation}\label{e.bae.Z}
Z_N^0(p_j) = \frac{2\pi I_j}{N} \quad\quad (j=1,\ldots,Q)
\end{equation}
where $I_j$ are half-integers (integers) for $Q$ even (odd).
If we define the energy $E_Q^R$ and $E_Q^L$ through the relation
\begin{equation} \label{lambtoe}
 \Lambda_{R,L}=\exp\left[v(N-2Q)-E_Q^{R,L}\right]\ ,
\end{equation}
they are given as
\begin{equation}\label{e.erl.def}
E_Q^{R,L} = -\left[ \pm \left(h+\frac{\delta}{2}\right)N+
\sum_{j=1}^Q \Phi^0_{R,L}(p_i)  \right]
\end{equation}
where
\begin{equation}\label{e.phirl}
\Phi^0_{R,L}(p)=\ln \frac{2\Delta-e^{\pm \delta}
 -e^{\pm ip}}{1-e^{\pm (\delta+i p)}},
\end{equation}
and $\{p_j\}$ are the distinct solutions of Eq.~(\ref{e.bae.Z}).
In Eqs.~(\ref{e.erl.def}) and (\ref{e.phirl}), the upper (lower) sign
corresponds to the R (L) case.

The bulk free energy is obtained from are ground-state energy. Using the
ground-state energy, the free energy $e$ as a function of $h$ and $q$ is
given as
\begin{equation}\label{bulk.fep}
{e}(h,q) = \lim_{N,Q\rightarrow\infty}
\min_{R,L} \left\{ \frac{E_Q^R}{N},\frac{E_Q^L}{N} \right\},
\end{equation}
where the limit is taken with fixed $q$.
The free energy $f$ as a function of $h$ and $v$ is given by the Legendre
transformation of $e$, i.e.,
\begin{equation}\label{Ff}
{f}(h,v) = \min_{0\le q \le 1} \{e(h,q) - (1-2q)v \}.
\end{equation}
The mean down-arrow density $\rho_D$ is the value of $q$ which minimizes
the expression above while the left-arrow density $\rho_L$ is given by
$\rho_L= \left(1+\partial e/\partial h\right)/2$.
Of the four model parameters, only $\Delta$ and $h$ enter the Bethe ansatz
equation determining $\{p_i\}$ while $\Phi^0_{R,L}$ depends
only on $\Delta$ and $\delta$. The vertical field $v$ simply adjusts the
mean value of $q$. Note that the horizontal field $h$ plays a role of an
additive constant to the phase function in Eq.~(\ref{e.Z0.def}).
This fact enables one to follow the same line of analysis as in~\cite{WoyET}.
The asymmetric XXZ Hamiltonian is diagonalized with the same Bethe ansatz
method but with a different energy function~\cite{Kim}. In next
section, we will show that the FSS properties do not depend on the form
of the energy function. So the asymmetric XXZ chain shares the same critical
property with the ASSV model.

Different choices of the set
$\{I_j\}$ in Eq.~(\ref{e.bae.Z}) lead to different eigenstates.
 It is well established that the
ground-state energy is obtained if $I_j= -(Q+1)/2+j$.
 In analogy with the free fermion theory, we will say that
a position $j$ is occupied by a particle (hole)
if $\phi_j \equiv -(Q+1)/2+j$
is included (not included)
 in the set $\{I_j\}$. Then the solution is classified by the
particle-hole configurations. The ground state corresponds to the Fermi sea
as shown in Fig.~2(a). An important class of levels is characterized by
shifting the Fermi sea by $m$, i.e., choosing $I_j=\phi_j+m$. We call this
the $m$-shifted levels and show in Fig.~2(b) an example.
Creating particles and holes at either ends of the $m$-shifted states
generates the whole class of excited states which scale
as $1/N$ in the critical phase.  General form of these excitations
is obtained by
creating $n_p$ particles at positions
$ j=Q+m+p_k $, ($k$=1,2,$\ldots$, $n_p$) with $1 \leq p_1 < p_2 < \cdots
 < p_{n_p}$ , $n_h$ holes at
$  j=Q+m+1-h_k$, ($k$=1,2,$\ldots$, $n_h$) with $ 1 \leq h_1 < h_2 < \cdots
 < h_{n_h}$,
 $\bar{n}_p$ particles at
$ j=1+m-\bar{p}_k $, ($k$=1,2,$\ldots$, $\bar{n}_p$)
 with $1 \leq \bar{p}_1 < \bar{p}_2 < \cdots
 < \bar{p}_{\bar{n}_p}$ , and $\bar{n}_h$ holes at
$ j=m+\bar{h}_k$, ($k$=1,2,$\ldots$, $\bar{n}_h$) with $ 1 \leq \bar{h}_1
 < \bar{h}_2 < \cdots
 < \bar{h}_{\bar{n}_h}$.  Without loss of generality we can set
$n_p=n_h$ and $\bar{n}_p=\bar{n}_h$. A particle-hole configuration is
collectively denoted by ${\cal P}$. And we will denote such an
excited level by
$(Q,m,{\cal P})$. In Figs.~2(c) and (d),
 we give some examples of particle-hole configurations.

For each range of $\Delta$, there exists a variable transformation
$p=p(\alpha)$ with which $\Theta^0(p(\alpha),p(\beta))$ defined in
Eq.~(\ref{e.theta.def}) depends only on $(\alpha-\beta)$~\cite{BukS,Nol,Bax}.
The resulting function is denoted by $\Theta$, i.e.,
$ \Theta(\alpha-\beta)\equiv \Theta^0(p(\alpha),p(\beta))$.
$Z_N(\alpha)$ and $\Phi_{R,L}(\alpha)$ are defined similarly. We use
superscript $^0$ for functions of $p$ and no superscript for functions of
$\alpha$.  In the thermodynamic limit $N\rightarrow\infty$ with fixed $q$,
the Bethe ansatz equation becomes an integral equation for the phase function
under the assumption that the solutions of Eq.~(\ref{e.bae.Z}) lie
densely on a smooth
curve ${\cal C}$ in the complex-$\alpha$ plane with end points $A$ and $B$.
For the ground state, the contour is symmetric with respect to the imaginary
axis so that $A=-B^*$.   Then $Z_\infty\equiv \lim_{N\rightarrow\infty}Z_N$
should satisfy the following integral equation
\begin{equation}\label{e.Z.int}
Z_\infty(\alpha) = p(\alpha) + 2hi - \frac{1}{2\pi}\int_{A}^{B}
\Theta(\alpha-\beta) Z'_\infty(\beta)\, d\beta \ .
\end{equation}
The solution of Eq.~(\ref{e.Z.int}) depends on $A$ and $B$ which are
determined as a function of $q$ and $h$ from the generalized normalization
condition
\begin{equation}  \label{normal1}
Z_\infty(B) = \pi q
\end{equation}
or, equivalently,
\begin{equation} \label{normal2}
Z_\infty(A)= -\pi q.
\end{equation}
Using the solution of Eq.~(\ref{e.Z.int}) and relation (\ref{normal1}), the
bulk free energy $e$ in Eq.~(\ref{bulk.fep}) is given by
\begin{equation}\label{bulk.fe}
e(h,q) = \min_{R,L} \left\{ \mp h \mp\frac{\delta}{2} -\frac{1}{2\pi}
\int_{A}^{B}\Phi_{R,L}(\alpha)Z'_\infty(\alpha)d\alpha \right\}
\end{equation}
where the upper (lower) sign corresponds to the R(L) case.

Before closing this section, we discuss the effect of the twisted  boundary
condition.
The boundary condition do not affect the bulk properties but change the
operator content and hence the TPF. The twisted boundary condition in the
context of the XXZ chain is to impose the condition $\sigma_{N+1}^{\pm} =
\exp(\pm 2\pi i l) \sigma_1^\pm$ where $\sigma_i^\pm$ are the Pauli spin
operators. In the vertex model language, it is equivalent to assign an
extra vertex weight $\exp (\pm \pi i l)$ to the horizontal arrow in the
first column. More generally, one can impose the twisted boundary conditions
$(l,l')$ by introducing the seams in the first
column and the first row where an extra weight $e^{-i\pi l}\ (e^{i\pi l})$
is assigned to each right (left) arrow in the first
column and similarly an extra weight $e^{i\pi l'}\ (e^{-i\pi l'})$ is
assigned to each  up (down) arrow in the first row.
Note that the effect of the vertical field is to give each up
(down) arrow on the lattice an extra weight $e^v$ ($e^{-v}$). However, since
the number of up arrows are conserved from row to row, one obtains the same
effect if a vertical field of strength $Mv$ is applied only to the first row
of vertical bonds.
Conversely, having
a seam with extra weights $e^{\pm i\pi l'}$ is equivalent to
assigning extra weights $e^{\pm i\pi l'/M}$ to all vertical bonds.
 A similar observation holds for the horizontal field also.
Therefore, one then sees that
the ASSV model with the twisted boundary conditions
$(l,l')$ is equivalent to the ASSV model with periodic boundary conditions
and with fields $\tilde{h}=h-\pi il/N$ and $\tilde{v}=v+\pi il'/M$.
In the XXZ chain, such relations are achieved by an unitary transformation
as discussed recently in~\cite{HenS}. A similar symmetry
operation exists
in the six-vertex model too~\cite{Noh}.

\section{Summation formula}\label{sec3}
In this section, we derive the summation formula which converts the sum of
the type
$$
{\cal S}[f^0] = \frac{1}{N} \sum_{j=1}^Q f^0(p_j)\ ,
$$
where $f^0$ is an arbitrary function which does not depend explicitly on $h$
and $q$ and $\{p_j\}$ is the solution of Eq.~(\ref{e.bae.Z})
for the $(Q,m,{\cal P})$ level, into a series in $1/N$.
We obtain the expressions for the expansion coefficients and derive the
useful relations between the correction terms and partial derivatives of the
bulk contribution of the sum.
The phase function in Eq.~(\ref{e.Z0.def})  and
the energy eigenvalue in Eq.~(\ref{e.erl.def}) involve such kind of sums.
For example,
\begin{equation}\label{fs.energy}
{E^{R,L}_Q} =-N\left[\pm\left(h+\frac{\delta}{2}\right)+ {\cal
S}[\Phi^0_{R,L}] \right].
\end{equation}

With the change of variable $p=p(\alpha)$
explained in Sec.~\ref{sec2}, the sum becomes
\begin{equation}\label{e.fs}
{\cal S}[f] = \frac{1}{N} \sum_{j=1}^Q f(\alpha_j)
\end{equation}
where $f(\alpha)\equiv f^0(p(\alpha))$ is not to be confused with the free
energy $f(h,v)$, $\alpha_j$ are given by
\begin{equation} \label{alphaj}
\alpha_j = Z_N^{-1} \left(\frac{2\pi I_j}{N}\right)
\end{equation}
with $\{I_j\}$ corresponding to the level $(Q,m,{\cal P})$, and
$Z_N^{-1}$ is the inverse function of $Z_N$.
Using Eq.~(\ref{alphaj}) in Eq.~(\ref{e.fs}) and making
explicitly reference to the locations of the
particles, one can re-write
Eq.~(\ref{e.fs})  as
\begin{eqnarray} \label{sumf}
{\cal S}[f] &=& \frac{1}{N} \sum_{j=1}^{Q} f\left(Z_N^{-1}(\phi_j)\right)
 \nonumber \\
       &&  + \frac{1}{N}
\sum_{j=1}^m \left[ f\left(Z_N^{-1}(\phi_{Q+j})\right) -
f\left(Z_N^{-1}(\phi_j)\right)\right] \nonumber \\
&&+ \frac{1}{N}\sum_{k=1}^{n_p}
\left[ f(Z_N^{-1}(\phi_{Q+m+p_k}))-f(Z_N^{-1}(\phi_{Q+m+1-h_k})) \right]
 \nonumber \\
&&+ \frac{1}{N}\sum_{k=1}^{\bar{n}_p} \left[
f(Z_N^{-1}(\phi_{1+m-\bar{p}_k}))-f(Z_N^{-1}(\phi_{m+\bar{h}_k})) \right]
\end{eqnarray}
where
\begin{equation}
\phi_j = -\frac{Q+1}{2} +j.
\end{equation}
The first sum is over the Fermi sea, the second accounts for the shift
of the Fermi sea, and the third (fourth) accounts for the particle-hole
configurations at the right (left) end of the Fermi sea.
The general strategy here is to regard $Z_N^{-1}$ as known first and
determine it self-consistently later. To proceed, we make the crucial
assumption that $Z_N^{-1}(\phi)$ has the first derivative at $\phi=\pm
\pi q$. We will see later that the $1/N$ scaling of the energy gaps, or the
mass gaps, throughout the critical phase
follows from this assumption. Conversely, we assume here that the
critical phase
is characterized by the fact that  $Z_N^{-1'} (\pm \pi q)$ exists with
possible exceptions at some special points such as at ($h$=0, $q$=1/2).
Note that  this assumption fails on the stochastic line
$\widetilde{\Delta}=1$ where the energy gap does not scale as
$1/N$ but as $1/N^{3/2}$~\cite{Kim}.
Now, applying the Euler-Maclaulin sum formula to the first sum of
 Eq.~(\ref{sumf})  and using the Taylor expansion of $f(Z_N^{-1}(\phi))$
at $\phi=\pm \pi q$ for the rest of the sums, one then obtains to $O(1/N^2)$,
\begin{eqnarray}\label{e.E_M_f}
{\cal S}[f] &=&
\frac{1}{2\pi} \int_{-\pi q}^{\pi q} f\left(Z_N^{-1}(\phi)\right)
d\phi +\left. \frac{m}{N}
 \left[f(Z_N^{-1}(\phi))\right]\right|_{-\pi q}^{\pi q}
\nonumber \\
&& + \frac{2\pi}{N^2}\left(\frac{m^2}{2}-\frac{1}{24}+{\cal N}\right)
f'\left(Z_N^{-1}(\pi q)\right)Z_N^{-1'} (\pi q) \nonumber \\
&&  - \frac{2\pi}{N^2}\left(\frac{m^2}{2}-\frac{1}{24}+
\overline{{\cal N}}\right)
f'\left(Z_N^{-1}(-\pi q)\right)Z_N^{-1'} (-\pi q) +
o\left(\frac{1}{N^2}\right)
\end{eqnarray}
where
\begin{eqnarray}
{\cal N} &=& \sum_{k=1}^{n_p} (p_k+h_k-1), \nonumber \\
\overline{\cal N} &=& \sum_{k=1}^{\bar{n}_p} (\bar{p}_k+\bar{h}_k-1).
\end{eqnarray}
This is not a correct ${1}/{N}$ expansion yet since $Z_N^{-1}(\phi)$ also has
$N$ dependence.
We next assume that $Z_N(\alpha)$ possesses the $1/N$ expansion of the form
\begin{equation}\label{e.ZN.exp}
Z_N(\alpha) = Z_\infty(\alpha) + \frac{b_1(\alpha)}{N} +
\frac{b_2(\alpha)}{N^2} +
o\left(\frac{1}{N^2}\right)
\end{equation}
where $Z_\infty$, $b_1$, and $b_2$ should be determined later.
Inverting Eq.~(\ref{e.ZN.exp}) using  the change of
variable $\phi = Z_\infty(\alpha)$,
the $N$ dependence of $Z_N^{-1}(\phi)$ can be shown to be
\begin{equation}\label{e.U.exp}
Z_N^{-1}(\phi) = \alpha - \frac{1}{N}\frac{b_1(\alpha)}{Z'_\infty(\alpha)} +
          \frac{1}{N^2}\left[ -\frac{b_2(\alpha)}{Z'_\infty(\alpha)} +
\frac{1}{Z'_\infty(\alpha)}\left(\frac{b_1^2(\alpha)}{2Z'_\infty(\alpha)}
\right)' \right] +o\left(\frac{1}{N^2}\right)
\end{equation}
where $'$ denotes the derivative with respect to $\alpha$. Inserting
Eq.~(\ref{e.U.exp}) into Eq.~(\ref{e.E_M_f}), expanding all quantities
to  $O(1/N^2)$ again, and changing the integration variable from
$\phi$ to $\alpha$, one finally obtains the summation formula in the form
\begin{mathletters}\label{e.fss.al}
\begin{equation}
{\cal S}[f] = {\cal S}_0 + \frac{{\cal S}_1}{N} +
\frac{{\cal S}_2}{N^2} + o\left(\frac{1}{N^2}\right)
\end{equation}
where
\begin{eqnarray}
{\cal S}_0 &=& \frac{1}{2\pi}\int_{A}^{B}f(\alpha)Z'_\infty(\alpha)d\alpha\ ,
\label{e.S0.res} \\
{\cal S}_1 &=& \frac{1}{2\pi}\int_{A}^{B}f'(\alpha)\left[2\pi
m-b_1(\alpha)\right] d\alpha \label{e.S1.res}\\
{\cal S}_2 &=& -\frac{1}{2\pi}\int_{A}^{B}f'(\alpha) b_2(\alpha) d\alpha
+ \frac{2\pi f'(B)}{Z'_\infty(B)} \left[ \frac{m^2}{2}
\left(1-\frac{b_1(B)}{2\pi m}\right)^2-\frac{1}{24}+{\cal N}\right]
\ ,\nonumber \\
&& - \frac{2\pi f'(A)}{Z'_\infty(A)} \left[ \frac{m^2}{2}
\left(1-\frac{b_1(A)}{2\pi m}\right)^2-\frac{1}{24}+\overline{\cal N}\right].
\label{e.S2.res}
\end{eqnarray}
\end{mathletters}
Here $A$ and $B=-A^*$ are determined from Eq.~(\ref{normal1}) or
Eq.~(\ref{normal2})  as functions
of $h$ and $q$.

In deriving Eq.~(\ref{e.fss.al}), we assumed the scaling form of $Z_N$ in
Eq.~(\ref{e.ZN.exp}). The self-consistent equations for $Z_\infty$, $b_1$,
and $b_2$ are obtained by applying the summation formula to
Eq.~(\ref{e.Z0.def}) with $f(\beta)=-\Theta(\alpha-\beta)$.
Equating corresponding orders in both sides of the resulting equation,
we obtain Eq.~(\ref{e.Z.int}) for $Z_\infty (\alpha)$ and
\begin{mathletters}\label{e.fss.Z}
\begin{eqnarray}
{\cal T}\circ[b_1(\alpha)] &=& m[\Theta(\alpha-A)-\Theta(\alpha-B)]
\label{e.b1.u} \\
{\cal T}\circ[b_2(\alpha)] &=& \frac{2\pi}{Z'_\infty(B)}
\left[ \frac{m^2}{2} \left(1-\frac{b_1(B)}{2\pi m}\right)^2-\frac{1}{24}+
{\cal N}\right] K(\alpha-B) \nonumber \\
&& -\frac{2\pi}{Z'_\infty(A)}
\left[\frac{m^2}{2} \left(1-\frac{b_1(A)}{2\pi m}\right)^2-
\frac{1}{24}+\overline{\cal N} \right]K(\alpha-A) \ , \label{e.b2.u}
\end{eqnarray}
\end{mathletters}
where $K(\gamma) \equiv d\Theta(\gamma)/d\gamma$ and ${\cal T}$ is a linear
operator defined by
\begin{equation}\label{Tdef}
{\cal T}\circ[G(\alpha)] \equiv G(\alpha) + \frac{1}{2\pi} \int_A^B
K(\alpha-\beta)G(\beta) \, d\beta
\end{equation}
for any function $G(\alpha)$.

The solutions of Eq.~(\ref{e.fss.Z}) are written in terms of the function
$F(\alpha,\mu,A,B)$~\cite{BogIK_BogIR} which is defined by
\begin{equation}\label{Fdef}
{\cal T}\circ[F(\alpha,\mu,A,B)] = -\frac{1}{2} \Theta(\alpha-\mu)\ .
\end{equation}
Because $\Theta$ is odd and $A=-B^*$, it satisfies the relation
\begin{equation}\label{Fprpt}
F(\alpha,\mu,A,B) = - F(-\alpha^*,-\mu^*,A,B)\ .
\end{equation}
Using the linearity of Eq.~(\ref{e.fss.Z}), $b_1(\alpha)$ can be written as
\begin{equation}\label{b1_D}
b_1(\alpha) = 2\pi m(1-D(\alpha))
\end{equation}
where
\begin{equation}\label{D_F}
D(\alpha)= 1-\frac{1}{\pi}\left[ F(\alpha,B,A,B)-F(\alpha,A,A,B)\right]\ ,
\end{equation}
the dressed charge function~\cite{BogIK_BogIR}.
Equation~(\ref{Fprpt}) implies that $D(\alpha) = D(-\alpha^*)$ and so
$D(A)=D(B)$. Using this, $b_2(\alpha)$ is written as
\begin{eqnarray}\label{d2_F}
b_2(\alpha) &=& -\frac{4\pi}{Z_\infty'(B)}
\left(\frac{D_0m^2}{2}-\frac{1}{24}+{\cal N}\right)
\stackrel{\circ}{F}(\alpha,B,A,B) \nonumber \\
&& +\frac{4\pi}{Z_\infty'(A)} \left(\frac{D_0m^2}{2}-\frac{1}{24}+
\overline{\cal N}\right) \stackrel{\circ}{F}(\alpha,A,A,B) \ ,
\end{eqnarray}
where $\stackrel{\circ}{F}(\alpha,\mu,A,B) \equiv -
\partial F(\alpha,\mu,A,B) /\partial \mu$ and
\begin{equation}\label{D0def}
D_0 \equiv D(A) = D(B)\ .
\end{equation}
Inserting these into Eqs.~(\ref{e.S1.res}) and (\ref{e.S2.res}), we can put
${\cal S}_1$ and ${\cal S}_2$ in the form
\begin{mathletters}\label{e.S12.sim}
\begin{eqnarray}
{\cal S}_1 &=&  m \int_{A}^B f'(\alpha) D(\alpha) d\alpha \ ,
\label{e.S1.sim} \\
{\cal S}_2 &=& 2\pi i\left(\frac{D_0^2m^2}{2}+{\cal N}-\frac{1}{24}\right)
\zeta
-2\pi i\left(\frac{D_0^2m^2}{2}+\overline{\cal N}-\frac{1}{24}\right)
\bar{\zeta} \ , \label{e.S2.sim}
\end{eqnarray}
\end{mathletters}
where
\begin{mathletters}\label{tau'.def}
\begin{eqnarray}
\zeta &=& \frac{1}{\pi i Z_\infty'(B)} \left( \int_{A}^B f'(\alpha)
\stackrel{\circ}{F}(\alpha,B,A,B) d\alpha
+ \pi f'(B)\right),\label{zeta} \\
\bar{\zeta} &=& \frac{1}{\pi iZ'_\infty(A) }\left( \int_{A}^B f'(\alpha)
\stackrel{\circ}{F}(\alpha,A,A,B) d\alpha
+ \pi f'(A)\right). \label{zetabar}
\end{eqnarray}
\end{mathletters}
We have used the notation $\zeta$ and $\bar{\zeta}$ for the quantities on
the right-hand side of Eqs.~(\ref{zeta}) and (\ref{zetabar}), respectively,
anticipating identification of them as the anisotropy factor. The fact
that $\bar{\zeta}$ is the complex conjugate of $\zeta$ is not transparent in
this form but will turn out to be the case as will be seen later.

The resulting expressions for ${\cal S}_1$ and ${\cal S}_2$ seem to be rather
complicated. But, the following manipulations show
 that they are related to the partial
derivatives of ${\cal S}_0$ with respective to $q$ and $h$.
To take necessary derivatives, one need to consider the variations of $A$
, $B$, and $Z_\infty(\alpha)$, which is denoted as $\delta A$, $\delta B$, and
$\delta Z_\infty (\alpha)$, respectively, upon the variations of $h$ and $q$,
denoted by $\delta h$ and $\delta q$, respectively.
We show in App. A that they are given by
\begin{mathletters} \label{delabz}
\begin{eqnarray}
Z_\infty'(A)\, \delta A &=&-(\pi + D_2 (A))\, \delta q-2iD_0\, \delta h\ ,\\
Z_\infty'(B)\, \delta B &=& (\pi - D_2 (B))\, \delta q-2iD_0\, \delta h\ ,\\
 \delta Z_\infty (\alpha)&=& D_2 (\alpha)\, \delta q + 2i D(\alpha)\,
  \delta h\ ,
\label{delzinfty}
\end{eqnarray}
\end{mathletters}
where
\begin{equation}\label{D2_F}
D_2(\alpha) \equiv F(\alpha,A,A,B)+F(\alpha,B,A,B) \ .
\end{equation}
{}From Eq.~(\ref{Fprpt}), one can see that $D_2(\alpha)=-D_2(-\alpha^*)$ which
implies that $D_2(A)=-D_2(B)$.
On the other hand, the variation of ${\cal S}_0$ given by
Eq.~(\ref{e.S0.res}) is
\begin{equation} \label{dels0}
\delta {\cal S}_0 = \frac{1}{2} (f(A)+f(B))\, \delta q - \frac{1}{2\pi}
 \int_A^B f'(\alpha)\, \delta Z_\infty (\alpha)\, d \alpha ,
\end{equation}
where a partial integration and Eqs.~(\ref{normal1}) and
(\ref{normal2}) are used. Combining
Eqs.~(\ref{dels0}) and (\ref{delzinfty}), one obtains
\begin{eqnarray}
\frac{\partial {\cal S}_0}{\partial h}&=& -\frac{i}{\pi}
  \int_A^B f'(\alpha) D(\alpha)\, d\alpha \label{s0h}\\
\frac{\partial {\cal S}_0}{\partial q} &=& \frac{1}{2} (f(A)+f(B)) -
\frac{1}{2\pi} \int_A^B f'(\alpha) D_2(\alpha)\, d\alpha\label{dS0dq}\
\end{eqnarray}
which, compared with Eq.~(\ref{e.S1.sim}), gives the first relation
\begin{equation}\label{S1_sus}
{\cal S}_1 = i \pi m \frac{\partial {\cal S}_0}{\partial h} \ .
\label{s1tosh} \end{equation}
It means that the first-order correction term is proportional to the
partial derivative of the bulk term with respect to $h$.
Next, the variations of $\partial {\cal S}_0/\partial h$ and $\partial
{\cal S}_0/\partial q$ are shown in App.~\ref{app0} to take the form
\begin{eqnarray}
\delta \left( \frac{\partial {\cal S}_0}{\partial h}\right) &=&
    J (\zeta+\bar{\zeta})\, \delta q - \frac{2iD_0^2}{\pi}
(\zeta-\bar{\zeta})\, \delta h\ , \label{delsh} \\
\delta \left( \frac{\partial {\cal S}_0}{\partial q}\right) &=&
\frac{i\pi J^2}{2D_0^2} (\zeta-\bar{\zeta})\, \delta q +
J(\zeta+\bar{\zeta})\, \delta h\ , \label{delsq}
\end{eqnarray}
where
$\zeta$, $\bar{\zeta}$ are given in Eqs.~(\ref{zeta}) and (\ref{zetabar}),
and
\begin{equation}\label{Jdef}
J \equiv D_0 (1-D_2(B)/\pi) =D_0 (1+D_2(A)/\pi)\ .
\end{equation}
$J$ is a constant depending on $h$, $q$, and $\Delta$ through
$A$, $B$, and $\Theta$.
But we find that the value of $J$ is equal to $1$ for any values of $A$,
and $B$ only if the function $\Theta$ is odd and $A=-B^*$.
We give the detail proof in App.~\ref{app2}.
Using the fact that $J=1$, we then have from Eq.~(\ref{delsh}) the
expression for $\zeta$ and
$\bar{\zeta}$ in terms of the partial derivatives of ${\cal S}_0$ as
\begin{equation} \label{zeta2}
\zeta = \frac{1}{2}\left[ \frac{\partial}{\partial q}\left( \frac{\partial
{\cal S}_0}{\partial h}\right) + \frac{\pi i}{2D_0^2}\frac{\partial^2 {\cal
S}_0}{\partial h^2}\right] =
        \frac{1}{2}\left[ \frac{\partial}{\partial h}\left( \frac{\partial
{\cal S}_0}{\partial q}\right) - \frac{2iD_0^2}{\pi}\frac{\partial^2 {\cal
S}_0}{\partial q^2}\right]
\end{equation}
and
\begin{equation} \label{zetabar2}
\bar{\zeta} = \frac{1}{2}\left[
 \frac{\partial}{\partial q}\left( \frac{\partial
{\cal S}_0}{\partial h}\right) - \frac{\pi i}{2D_0^2}\frac{\partial^2 {\cal
S}_0}{\partial h^2}\right] =
        \frac{1}{2}\left[ \frac{\partial}{\partial h}\left( \frac{\partial
{\cal S}_0}{\partial q}\right) + \frac{2iD_0^2}{\pi}\frac{\partial^2 {\cal
S}_0}{\partial q^2}\right] .
\end{equation}
 Since ${\cal S}_0$ is a real function of $h$ and $q$, it is now
obvious that $\bar{\zeta}$ is indeed the complex conjugate of $\zeta$ as
claimed before. From the second equalities in Eqs.~(\ref{zeta2}) and
(\ref{zetabar2}), we find the identity
\begin{equation} \label{idnt_p}
- \frac{\frac{\partial^2 {\cal S}_0}{\partial q^2}}{\frac{\partial^2
{\cal S}_0}{\partial h^2}} = \left( \frac{\pi}{2D_0^2} \right)^2 .
\end{equation}
This identity comes from the fact that $J=1$ and is very important in
identifying the universality class of the critical phase of the ASSV model.
Equations~(\ref{s1tosh}),
(\ref{zeta2}), and (\ref{zetabar2}) together with Eq. (\ref{e.S2.sim})
relate the finite size corrections to the derivatives of ${\cal S}_0$.
An analogous relation between the Fermi velocity and magnetic
susceptibility has been obtained by Bogoliubov {\it et al}.~\cite{BogIK_BogIR}
for the symmetric XXZ chain but our result is more general.

\section{Finite-size corrections of the transfer-matrix spectra}\label{sec4}
In this section, we investigate the finite-size corrections to the transfer
matrix spectra of the ASSV model using the summation formula obtained in
Sec.~\ref{sec3}.
For simplicity we only mention the case where $E_Q^R < E_Q^L$ which we call
the R case. Most of results are the same for both  R and L cases.
The energy for the level $(Q,m,{\cal P})$ is obtained by applying the
summation formula to Eq.~(\ref{fs.energy}):
\begin{equation}\label{FSS.ENERGY}
E_{Q,m,{\cal P}} = -\left(h+\frac{\delta}{2}+{\cal S}_0[\Phi_R]\right) N
-{\cal S}_1[\Phi_R]-\frac{{\cal S}_2[\Phi_R]}{N}
 + o\left(\frac{1}{N}\right) \ ,
\end{equation}
where the leading order term contributes to the bulk free energy, i.e.,
${\cal S}_0 = -e-h-\delta/2$ from Eq.~(\ref{bulk.fep}). The correction term
${\cal S}_1$ is obtained from Eq.~(\ref{s1tosh})
\begin{equation}\label{S1.od}
{\cal S}_1 = -\pi im(e_h+1)
\end{equation}
and ${\cal S}_2$ is given by Eq.~(\ref{e.S2.sim}) with
\begin{equation}\label{tau'_def}
\zeta =- \frac{1}{2}e_{h,q} - \frac{\pi i}{4D_0^2}e_{h,h} =
-\frac{1}{2}e_{q,h} + \frac{iD_0^2}{\pi} e_{q,q}
\end{equation}
from Eq.~(\ref{zeta2}).
We use the short-hand
notations that the subscripts of $e$ denote partial derivatives.
Note that $e_h=2\rho_L-1$ where $\rho_L$ is the left-arrow
density (For the L case, $(e_h+1)$ in Eq.~(\ref{S1.od}) is replaced by
$(e_h-1)$).

The partition function is
\begin{equation}   \label{partftn}
{\cal Z}= \sum_{Q,m,{\cal P}} e^{-M E_{Q,m,{\cal P}}+vM(N-2Q)}
\end{equation}
with $E_{Q,m,{\cal P}}$ given in Eq.~(\ref{FSS.ENERGY}). To sum over the
sectors $Q$, we take advantage of the fact that  the summand
in Eq.~(\ref{partftn}) is peaked around the value of $Q$ near $\overline{Q}
\equiv N \rho_D$.
Inserting $q=\rho_D+(Q-\overline{Q})/N$ to Eq.~(\ref{FSS.ENERGY}) with
a term $-v(1-2q)N$ added and expanding to order ${1}/{N^2}$ assuming
$Q-\overline{Q}$ is of $O(1)$, one gets the scaling form of the energy as
\begin{eqnarray}\label{e.fss.a6v}
E_{Q,m,{\cal P}}-v (1-2q)N &=& f(h,v) N + 2\pi im\rho_L
-\frac{2\pi i\zeta '}{N}\left(m(Q-\overline{Q})+{\cal N}-\overline{\cal
N}\right) \nonumber \\
&&+\frac{2\pi\zeta''}{N}\left(D_0^2m^2 +\frac{e_{q,q}}{4\pi
\zeta ''}(Q-\overline{Q})^2+{\cal N}+\overline{\cal N}-\frac{1}{12}\right)
+o\left(\frac{1}{N}\right)
\end{eqnarray}
where $f(h,v)$ is given in Eq.~(\ref{Ff}), $\zeta '$ $(\zeta '')$ is the
real (imaginary) part of $\zeta$ given in Eq.~(\ref{tau'_def}),
and the relation ${\partial \rho_L}/{\partial q}=e_{h,q}/2$
has been used.

On the other hand, for models whose TPF is given by Eq.~(\ref{coulomb}),
the $O(1/N)$ part of the energy is expected to behave as
$$ \frac{2\pi \zeta ''}{N} \left(\frac{m^2}{2g}+\frac{gn^2}{2} +{\cal
N}+\overline{\cal N}-\frac{1}{12}\right)
-\frac{2\pi i\zeta '}{N}\left(mn+{\cal N}-\overline{\cal
N}\right). $$
Comparing Eq.~(\ref{e.fss.a6v}) with this expression
and identifying the indices $m$ and $Q-\bar{Q}$
in  Eq.~(\ref{e.fss.a6v}) as the spin-wave and vortex
indices, respectively, as in the SSV model,
 one sees  that
the critical phase of the ASSV model is indeed in the
Gaussian-model universality class with $c=1$.
The Gaussian coupling constant $g$ can be
read off from the coefficient of $m^2$ in Eq.~(\ref{e.fss.a6v})
as
\begin{equation}\label{e.rst.g}
g = \frac{1}{2D_0^2}
\end{equation}
where we recall that $D_0$ is defined in Eqs.~(\ref{D0def}) and (\ref{D_F}).
However, from the coefficient of $(Q-\bar{Q})^2$
in Eq.~(\ref{e.fss.a6v}), we also have the relation
\begin{equation}  \label{e.rst.g2}
g= \frac{e_{q,q}}{2\pi \zeta ''}.
\end{equation}
Therefore, to identify the spectra Eq.~(\ref{e.fss.a6v}) with those of the
Gaussian model, we require that the two expressions of $g$,
Eqs.~(\ref{e.rst.g}) and (\ref{e.rst.g2}), give the identical result.
Using Eqs.~(\ref{tau'_def}) and (\ref{e.rst.g}), the condition is then
\begin{equation}\label{e.ratio.univ}
-\frac{e_{q,q}}{e_{h,h}} = (\pi g)^2
\end{equation}
which is guaranteed by the identity (\ref{idnt_p}). That the coupling
constant $g$ is given by Eq.~(\ref{e.rst.g}) is first derived by Izergin and
Korepin~\cite{IzeK} for the symmetric XXZ chain in a magnetic field but it
also holds for the ASSV model.  Using the relations
\begin{eqnarray*}
 e_{h,h} &=& \frac{f_{h,h}f_{v,v}-f_{h,v}f_{v,h}}{f_{v,v}} \\
 e_{q,q} &=& -\frac{4}{f_{v,v}} \ .
\end{eqnarray*}
in Eq.~(\ref{e.ratio.univ}), we obtain an alternative expression given in
Eq.~(\ref{hessian}).

Before proceeding further, we should ask the validity of the summation
formula. If the summation formula is valid with non-trivial solution for
$b_2$, i.e., $b_2(\alpha)\neq 0$, then transfer matrix spectra follow the
scaling form in Eq.~(\ref{e.fss.crt}) which holds only in critical phase.
We have assumed the expansion forms of $Z_N$ and $Z_N^{-1}$ in
Eqs.~(\ref{e.ZN.exp}) and (\ref{e.U.exp}) in deriving the summation formula.
The assumption is valid only when the coefficients are finite, i.e.,
$Z'_\infty$ should not be $0$. Otherwise, the summation formula does not hold
and the criticality is not guaranteed.
So it is the sufficient condition for the criticality that $Z_\infty'(\alpha)$
and $b_2(\alpha)$ are not zero. It is compatible with earlier studies on the
phase diagram of the asymmetric six-vertex model.
First, for $\Delta\ge 1$ $Z'_\infty(A)$ is always $0$ for any values of $q$
when $\widetilde{\Delta}=1$~\cite{BukS}.
At this point, the ASSV model describes stochastic
growth models and the finite-size corrections show the
Kardar-Parisi-Zhang scaling~\cite{GwaS,Kim}. Another case occurs when
$\Delta<-1$. In this region, there exists the anti-ferroelectric
phase when $q=1/2$. The phase boundary is given in~\cite{Nol} and
$Z'_\infty(A)$ at the phase boundary is also $0$. The finite-size corrections
on this phase boundary has recently been studied by Albertini {\it et
al.}~\cite{Albertini}.
There are also ordered ferroelectric phase with $q=0$ in which case
$A=B$ from Eqs.~(\ref{normal1}) and (\ref{normal2}).
In this case $b_2(\alpha)$ is identically zero and the system is also out of
criticality. The $q=0$ and $1/2$ ordered phases are separated from the critical
phase by the Pokrovsky--Talapov (PT)  transition~\cite{LieW}.

Inside the phase boundaries stated above, the system is critical and
the finite-size corrections to the transfer matrix spectra are obtained from
Eqs.~(\ref{FSS.ENERGY}), (\ref{S1.od}), (\ref{e.S2.sim}), and
(\ref{tau'_def}).
Actual values of $g$ through the critical phase can be mapped out without
difficulty.
In Figs.~3, we present the constant-$g$ lines in the $(\widetilde{\Delta},q)$
plane for $\tanh(2h)=0$, $0.5$, and $1.0$.
As seen from Eq.~(\ref{e.rst.g}), $g$ is obtained from the function $D$.
Instead of solving Eq.~(\ref{Fdef}) for $F$ and using Eq.~(\ref{D_F}), we
study the integral equation for $D(\alpha)$
\begin{equation}\label{Dint}
{\cal T}\circ[D(\alpha)] = 1\ ,
\end{equation}
which is obtained by combining Eqs.~(\ref{D_F}) and (\ref{Fdef}). It
can be solved analytically only for special cases.
For $q=0$, there is a trivial solution $D_0(\alpha)=1$.   So we have
$g=1/2$ at $q=0$ and at $q=1$ by symmetry.
For $\Delta(=-\cosh\lambda)<-1$,  the anti-ferroelectrically ordered phase
with $q=1/2$ appears for $\widetilde{\Delta}<\widetilde{\Delta}_c$ with
\begin{equation}\label{delta_c}
\widetilde{\Delta}_c \equiv \frac{\Delta}{\cosh(2h_c)}
\end{equation}
where $h_c$ is given by~\cite{Nol}
\begin{equation}
h_c = \ln{\cosh{\lambda}} - \frac{\lambda}{2}-\sum_{n=1}^\infty
\frac{(-1)^n}{n} e^{-2n\lambda}\tanh(\lambda n)\ .
\end{equation}
The value of $ \widetilde{\Delta}_c$ decrease from $-1$ for $\tanh(2h)=0$
to $-4$ for $\tanh(2h)=\pm 1$.
At the phase boundary, the value of $B$ is $\pi+i\lambda$~\cite{Nol}.
Then Eq.~(\ref{Dint}) is solved by Fourier series
method to yield the solution $D(\alpha)=1/2$. So we have
$g=2$ at the anti-ferroelectric
phase boundary. For $\Delta > 1$, the ASSV model is critical only for
$\widetilde{\Delta}<1$. As $\widetilde{\Delta}$
approaches 1 from below, $g$ decreases to 0.
For the SSV model, the model is critical for $-1< \Delta (=
 -\cos \gamma) < 1$ and the value of $g$ is given by the simple formula
$g=1-\gamma/\pi$~\cite{Alcaraz}.
In this case, $B\rightarrow \infty$ and naive application
of Eq.~(\ref{Dint}) is invalid due to the fact that $Z_\infty'(\pm \infty)=0$.
However, a careful treatment using the
Wiener-Hopf method can reproduce the correct result starting from
Eq.~(\ref{Dint})~\cite{BogIK_BogIR,Noh}.
For other cases, the values of $g$ are obtained by solving the integral
equation (\ref{Dint}) numerically. Note that there is a discontinuity in
$g$ from $1$ to $2$ at $(\widetilde{\Delta},q)=(-1,1/2)$ for $\tanh(2h)=0$.
It is the Kosteritz-Thouless~(KT)
 transition point where the free energy has the
essential singularity~\cite{Bax}. There is a crossover from the KT transition
to the PT transition for $\tanh(2h)\neq 0$.

We next derive the TPF $\widetilde{\cal Z}$.
Using the scaling form of Eq.~(\ref{e.fss.a6v}) and the identity
Eq.~(\ref{e.ratio.univ}), it can be put in the form
\begin{equation}\label{e.mpf}
\widetilde{\cal Z} = \sum_{Q,m\in {\bf Z}}\sum_{\cal P} e^{-2\pi im\rho_L M}
\exp\left[ 2\pi i \tau \left(\Delta_{m,Q-\overline{Q}}+{\cal N} -
\frac{1}{24} \right) - 2\pi i \bar{\tau} \left(
\overline{\Delta}_{m,Q-\overline{Q}}+\overline{\cal N}-\frac{1}{24}
\right)  \right]
\end{equation}
where the modular ratio  $\tau$ is given as
$\tau = \frac{M}{N}\zeta$.
The sum over ${\cal P}$ produces the factor
$$
\sum_{\cal P} {\sf q}^{{\cal N}-\frac{1}{24}}
 \bar{\sf q}^{\overline{\cal N}-\frac{1}{24}}
= \frac{1}{\eta({\sf q})\eta(\bar{\sf q})}
$$
where the nome
 ${\sf q}=e^{2\pi i\tau}$ and $\eta({\sf q})$ is the Dedekind eta function
given in Eq.~(\ref{dedekind}).
Now we introduce the two mismatch parameters
\begin{equation} \label{mismatch}
\alpha=\{\rho_L M\} \makebox[2cm]{and} \beta=\{\rho_D N\}=\{\overline{Q}\}
\end{equation}
where $\{x\}$ denotes the fractional part of $x$; i.e., $\alpha$ and $\beta$
are the mean number of left arrows per column and down arrows per row,
respectively, modulo 1.  We next change the index $Q$ in Eq.~(\ref{e.mpf})
to $n$ by the relation $Q-\bar{Q}=n-\beta$.
Then the TPF for the critical phase of the ASSV model is given in the form
of the modified Coulombic partition function as given in
Eq.~(\ref{Z_mpf.a6v}).
The mismatches
come from the incommensuration of arrow densities $\rho_L$ and $\rho_D
$ with system
size $M$ and $N$, respectively.

We also consider the effect of the twisted boundary
condition on the TPF. Since the twisted boundary  conditions $(l,l')$
corresponds to
the ASSV model with the periodic boundary conditions but with the
modified fields $h-\pi il/N$
and $v+\pi il'/M$, these $O(1/N)$ changes of $h$ and $v$ modifies
the TPF. Thus, one replaces $h$ and $v$ by $h-\pi il/N$
and $v+\pi il'/M$, respectively, in Eq.~(\ref{e.fss.a6v}),
and makes Taylor expansion to the necessary order in $1/N$ to obtain the
boundary
condition effect on the finite-size corrections. After a straightforward
calculation the resulting TPF is found to be generalized to
\begin{equation}\label{mpf_tw}
\widetilde{\cal Z}(\rho_L,\rho_D;l,l';M,N|\tau) =
e^{\pi iMl(2\rho_L-1)-\pi iNl'(2\rho_D-1)} \widetilde{\cal Z}_{\text{gC}}
(\{M\rho_L\},\{N\rho_D\}; l,l'|\tau)
\end{equation}
where $\widetilde{\cal Z}_{\text{gC}}$ is the generalized Coulombic
partition function defined as
\begin{equation}\label{mpf_p}
\widetilde{\cal Z}_{\text{gC}}(\alpha,\beta;l,l'|\tau) \equiv
\frac{1}{|\eta({\sf q})|^2} \sum_{m,n\in {\bf Z}}
e^{-2\pi i m\alpha -2\pi il'(n-\beta)}\ \
{\sf q}^{\Delta_{{m-l},n-\beta}}\bar{\sf q}^{
\overline{\Delta}_{{m-l},{n-\beta}}}\ .
\end{equation}
The toroidal partition functions in Eqs.~(\ref{Z_mpf.a6v}) and (\ref{mpf_tw})
satisfy the necessary modular covariance as they should be. These are
discussed in App.~\ref{app3}.

\section{Discussions}\label{sec5}

In this paper, we utilized a method of calculating the finite-size
corrections applicable to the critical phase of the ASSV model
and investigated the FSS of the transfer-matrix spectra of the model for
arbitrary sector $Q=qN$. Since we work for general $q$, the string solutions
appearing only at $q=1/2$ do not complicate the analysis~\cite{YuF} and
low-lying excitations are easily classified. For any low-lying levels, the
finite-size corrections for arbitrary sum over functions of rapidities are
related to partial derivatives of the bulk contributions; Eq.~(\ref{S1_sus})
for the first order term and Eqs~(\ref{e.S2.sim}), (\ref{zeta2}), and
(\ref{zetabar2}) for the second order term.
{}From this, finite-size scaling amplitudes of the ASSV model are related
to thermodynamic quantities.

The ASSV model is conveniently parameterized by the interaction parameter
$\widetilde{\Delta}$ (Eq.~(\ref{delta_tilde})), the anisotropy $\delta$
(Eq.~(\ref{energy_assign})), the horizontal field $h$ and finally vertical
field $v$ or alternatively, the down arrow density $q$. For fixed $h$,
$\delta$, and $q(\neq1/2)$, the model is critical for
$-\infty<\widetilde{\Delta}<1$. When $q=1/2$, however, the region defined by
$\widetilde{\Delta}<\widetilde{\Delta}_c$ with $\widetilde{\Delta}_c$ given by
Eq.~(\ref{delta_c}) corresponds to the anti-ferroelectrically ordered phase.
We showed that the
critical phase is in the Gaussian-model universality class with $c=1$ and
the Gaussian coupling constant $g$ as
given by Eq.~(\ref{e.rst.g}). The value of $g$ ranges from 0 at the
stochastic limit $\widetilde{\Delta}=1$ to 1/2 at the PT transition lines
bordering the $q=0$ or $q=1$ phases and to 2 at the PT transition lines
bordering the $q=1/2$ anti-ferroelectrically ordered phase.

We also constructed the TPF in Eq.~(\ref{Z_mpf.a6v}) for periodic boundary
conditions and in
Eq.~(\ref{mpf_tw}) for more general twisted boundary conditions.
 Due to the incommensuration of arrow densities, it takes
an additional phase factor $e^{-2\pi im\alpha}$ and
a constant shift $\beta$ in the
vortex-excitation index $n$. The phase factor comes from the $O(1)$
imaginary term of the energy spectra Eq.~(\ref{S1.od}) which can be
understood as follows.
The continuous lines connecting the left pointing arrows may be pictured as
domain walls or steps running across the lattice. Since there are $\rho_L M$
of them, the mean distance between them is $1/\rho_L$ lattice units.
Thus the periodicity of the lattice in the time direction is enlarged by the
same factor and the transfer matrix ${\bf T}$ is the $\rho_L$-th power
of the time translation operator for one unit distance. Therefore, a factor
$\exp ( 2\pi i \rho_L)$ may appear in the spectrum.
As shown in App.~\ref{app3}, the  mismatch parameter in the phase factor
of Eq.~(\ref{Z_mpf.a6v}) becomes that which shifts the vortex-excitation
index upon
the modular transformation which simply rotates the lattice by 90$^0$
exchanging the role of $M$ and $N$.

Our result for the TPF is consistent with all other previous results.
For the five-vertex model in the noninteracting case, the TPF on a deformed
square lattice is given as~\cite{NK1}
\begin{equation}\label{5v}
\widetilde{\cal Z}_{\text{5-v}} = \frac{1}{|\eta|^2} \sum_{m,n\in {\bf Z}}
e^{-\pi i Q_1 m} {\sf q}^{\Delta_{m,n-Q_0}}\bar{\sf q}^{\overline{
\Delta}_{m,n-Q_0}}\ ,
\end{equation}
where $Q_0=(1-\rho_D)N$ and $Q_1=(2\rho_L-\rho_D-1)M$.
The TFP's on different geometry can be transformed to each other through the
modular transformation~[See App.~\ref{app3}]. One can find that Eq.~(\ref{5v})
is obtained from Eq.~(\ref{Z_mpf.a6v}) through the modular
transformation $\tau\rightarrow \tau+M/(2N)$ with $M/(2N)=1$.
In the SSV model, $\rho_L$=$\rho_D$=1/2 and hence $\alpha$ ($\beta$) is
0 if $M$ ($N$) is even and 1/2 if odd. If the twisted boundary condition
$(l,0)$ is applied in the SSV model,
the operator content of this model obtained by other methods
is summarized by the TPF~\cite{KimP2}
\begin{equation}
\widetilde{\cal Z} =
\frac{1}{|\eta({\sf q})|^2} \sum_{m,n\in{\bf Z}} e^{2\pi im\mu}
{\sf q}^{\Delta_{m-l,n-\nu}}\bar{\sf q}^{\Delta_{m-l,n-\nu}}
\end{equation}
where $\mu(\nu)$ is $0$ for $M(N)$ even and $1/2$ for odd.
It is in agreement with Eq.~(\ref{mpf_tw}) when one uses
$\rho_L=\rho_D=\frac{1}{2}$ and $l'=0$ in  Eq.~(\ref{mpf_tw}).

The asymmetric XXZ chain is obtained if one takes an extreme-anisotropic
limit in which the vertex-weights approaches the limit
$w_3=w_4=0$ and $w_1=w_2=w_5=w_6=1$. With an appropriate parameterization,
${\bf T}$ can be put in the form
\begin{equation}
{\bf T} (u)= \exp ( i {\bf P}-u {\bf H}_{\text{XXZ}} + O(u^2) )
\end{equation}
where $u$ is the so-called spectral parameter as a function of which ${\bf
T}(u)$ form a commuting family, ${\bf P}$ is the shift operator,
and ${\bf H}_{\text{XXZ}}$ is the asymmetric XXZ chain Hamiltonian.
Therefore, ${\bf T}$ and  ${\bf H}_{\text{XXZ}}$ share a same set of
spectra. However, the TPF is not the same as  Tr $\exp (- u M {\bf
H}_{\text{XXZ}} )$. This is why complete information on
the operator content for the XXZ chains is not sufficient to construct the
TPF of the lattice model.

When the ASSV model is considered as a model for the ECS of e.g.
fcc (110) surface, the free energy $f(h,v)$ itself is the height of the
surface from the base (110) plane with appropriate identification of
coordinates.  In particular, the surface curvature $\kappa$ defined in
e.g. \cite{ecs} is related to the Hessian of $f$ by
\begin{equation}  \label{curvature}
\kappa = \frac{d^2}{k_B T} H^{1/2}
\end{equation}
where $d$ is the distance between the crystal planes. Our result
Eq.~(\ref{hessian}) then relates $\kappa$ to $g$ by
\begin{equation} \label{kappatog}
\kappa = \frac{2 d^2}{\pi  k_B T} \frac{1}{g}
\end{equation}
exactly. When the interactions are intrinsically antiferroelectric,
 the $q=1/2$ phase at low temperatures corresponds to the flat (110) facet.
As the temperature is raised the facet area becomes smaller and finally
disappears at the roughening temperature $T_R$. The roughening transition
corresponds to the KT point where $g=1$. Thus there occurs a universal jump
in the curvature of magnitude $2d^2/(\pi k_B T_R)$.
This universal jump has been anticipated from the
Coulomb gas picture of the solid-on-solid models~\cite{villain}
and also from the solution
of the six-vertex model near $h$=$v$=0~\cite{ecs} and has been
measured experimentally~\cite{gallet}.
Our result goes beyond this. Equation~(\ref{kappatog}) being exact,
it applies to all points of the curved portion of the crystal.
In particular, near the PT transitions where the curved surface joins
the facet smoothly with the exponent $\theta=3/2$~\cite{rottman}
, one expects the universal jump of the curvature
with $g=2$ or $g=1/2$ depending on which portion of the PT line is
appropriate. It is interesting to speculate whether the relation
Eq.~(\ref{kappatog}) also holds for other solid-on-solid type models and for
real samples. If it is valid for real systems, it would provide a mean to
measure the critical exponent directly from surface curvatures.

\acknowledgements

This work is supported by KOSEF through the Center for Theoretical Physics,
SNU, by the Ministry of Education grant BSRI-94-2420, and also by NSF
grant DMR-9205125. DK thanks Prof. M. den Nijs for hospitality and discussions
during his visit to University of Washington where most of this work is carried
out.

\begin{appendix}
\section{Variations of the partial derivatives of ${\cal S}_0$}\label{app0}

In this appendix, we consider the variations of ${\cal S}_0$ upon the
variations of $h$ and $q$. As $h$ and $q$ vary, the end points $A$ and $B$
also vary satisfying
\begin{equation} \label{ap1}
\pi\, \delta q = \delta Z_\infty (B) + Z_\infty '(B)\, \delta B
\end{equation}
and
\begin{equation}  \label{ap2}
-\pi\, \delta q = \delta Z_\infty (A) + Z_\infty '(A)\, \delta A \ .
\end{equation}
They are obtained from the variations of Eqs.~(\ref{normal1}) and
(\ref{normal2}). Since this involves $\delta Z_\infty (\alpha)$, we
take the variation of Eq.~(\ref{e.Z.int}) after performing the
partial integration inside the integral operator. This gives a simpler
result
\begin{equation}   \label{ap3}
\delta Z_\infty (\alpha) = 2i D(\alpha)\, \delta h + D_2 (\alpha)\, \delta q
\end{equation}
with $D$ and $D_2$ defined by Eqs.~(\ref{D_F}) and (\ref{D2_F}).
In taking the variations in this appendix, it is convenient to use the
integral representations of $D$ and $D_2$ rather than
the definitions in Eqs.~(\ref{D_F}) and (\ref{D2_F}).
The integral representation for $D$ is given in Eq.~(\ref{Dint}) and the other
for $D_2$ are easily obtained by combining Eqs.~(\ref{D2_F}) and (\ref{Fdef}):
\begin{equation}
{\cal T}\circ[D_2(\alpha)] = -\frac{1}{2}[\Theta(\alpha-A)+\Theta(\alpha-B)]\
,\label{D2.intp}
\end{equation}
where the operator ${\cal T}$ is defined in Sec.~\ref{sec3}.
Inserting Eq.~(\ref{ap3}) into Eqs.~(\ref{ap1}) and (\ref{ap2}) gives
Eq.~(\ref{delabz}).

Next, consider the variation of Eqs.~(\ref{s0h}) and (\ref{dS0dq}):
\begin{eqnarray}
\delta \left( \frac{\partial {\cal S}_0}{\partial h}\right) &=&
-\frac{i}{\pi} \left( \int_A^B f'(\alpha) \delta D(\alpha) d\alpha
 + f'(B) D_0 \delta B - f'(A) D_0 \delta A \right)
\label{ap4}\ , \\
\delta \left( \frac{\partial {\cal S}_0}{\partial q}\right) &=&
-\frac{1}{2\pi}\int_A^B f'(\alpha)\, \delta D_2(\alpha)\, d\alpha \nonumber \\
&&+ \frac{f'(A)}{2}\left(1+\frac{D_2(A)}{\pi}\right)\, \delta A +
\frac{f'(B)}{2}\left(1-\frac{D_2(B)}{\pi}\right)\, \delta B\ .\label{dels0q}
\end{eqnarray}
The variations of $D(\alpha)$ and $D_2(\alpha)$ are obtained from
Eqs.~(\ref{Dint}) and (\ref{D2.intp}). After a straightforward calculation,
one can find that
\begin{eqnarray}
\delta D(\alpha) &=& \frac{D_0}{\pi} \left\{\stackrel{\circ}{F}(\alpha,B,A,B)
\, \delta B - \stackrel{\circ}{F}(\alpha,A,A,B)\,  \delta A \right\}
\label{ap5}\ , \\
\delta D_2(\alpha) &=& -\left(1-\frac{D_2(B)}{\pi}\right)
\stackrel{\circ}{F}(\alpha,B,A,B)\, \delta B -
\left(1+\frac{D_2(A)}{\pi}\right)
\stackrel{\circ}{F}(\alpha,A,A,B)\, \delta A\ . \label{delD2}
\end{eqnarray}
Using Eqs.~(\ref{ap5}) and (\ref{delD2}) in Eqs.~(\ref{ap4}) and
(\ref{dels0q}) and using the definitions of
$\zeta$ and $\bar{\zeta}$  in Eq.~(\ref{tau'.def}), one obtains
\begin{eqnarray}
\delta \left( \frac{\partial {\cal S}_0}{\partial h}\right) &=&
-\frac{iD_0}{\pi} \left( i\zeta  Z_\infty'(B)\, \delta B
-i\bar{\zeta}  Z_\infty'(A)\,  \delta A \right) \label{ap6} \\
\delta \left( \frac{\partial {\cal S}_0}{\partial q}\right) &=&
\frac{1}{2}\left(1-\frac{D_2(B)}{\pi}\right) i\zeta Z'_\infty(B)\, \delta B
+\frac{1}{2}\left(1+\frac{D_2(A)}{\pi}\right) i\bar{\zeta}Z'_\infty(A)\,
\delta A
\label{dels0q_m}
\end{eqnarray}
which together with Eq.~(\ref{delabz}) give Eqs.~(\ref{delsh}) and
(\ref{delsq}).

\section{The proof of $J=1$}\label{app2}
In this appendix we present some identities between the partial derivatives
of $F(\alpha,\mu,A,B)$ defined in Eq.~(\ref{Fdef}) and prove the identity
$J\equiv D(B)(1-D_2(B)/\pi)=1$, where
\begin{eqnarray}
D(B) &=&  1-\frac{1}{\pi}[F(B,B,A,B)-F(B,A,A,B)]\ , \\
D_2(B) &=& F(B,B,A,B)+F(B,A,A,B)\ .
\end{eqnarray}
In general, $J$ may be a function of $a$ and $b$ with $A=-a+ib$ and $B=a+ib$.
But, Eq.~(\ref{Fdef}) is invariant under the shift of all arguments
by an imaginary amount, i.e.,
$F(\alpha,\mu,A,B)=F(\alpha+iu,\mu+iu,A+iu,B+iu)$ for any real $u$. This means
that $D(B)$, $D_2(B)$, and thus $J$ are functions of only $a$.
When $a=0$ the function $F(\alpha,\mu,A,B)$ becomes simplify
$-{{1}\over{2}}\Theta(\alpha-\mu)$ and so $J=1$.
Thus, for the proof of $J=1$, it suffices to show that the total derivative
of $J$ with respect to $a$ is identically $0$.

The total derivative of $J$ contains the partial derivatives of $F(B,B,A,B)$
and $F(B,A,A,B)$ with respect to $a$:
\begin{eqnarray}
\frac{d}{da}F(B,B,A,B) &=& F'-\stackrel{\circ}{F}-F_A+F_B\ ,\label{F1_a} \\
\frac{d}{da}F(B,A,A,B) &=& F'+\stackrel{\circ}{F}-F_A+F_B\ ,\label{F2_a}
\end{eqnarray}
where $F'\equiv \partial F/\partial \alpha$, $\stackrel{\circ}{F}\equiv
-\partial F/\partial \mu$, $F_A\equiv \partial F/\partial A$,
and $F_B\equiv \partial F/\partial B$. We use the convention that the
arguments of the functions are $(\alpha,\mu,A,B)$ when not shown explicitly.
First, we derive several identities between partial derivatives of $F$, which
will simplify Eqs.~(\ref{F1_a}) and (\ref{F2_a}).
Taking partial derivatives of Eq.~(\ref{Fdef}), one obtains integral equations
for $F'$, $F_A$, and $F_B$. After some manipulations, they take the form
\begin{eqnarray}
{\cal T}\circ F' &=& -\frac{1}{2}K(\alpha-\mu)
                 + \frac{1}{2\pi}\left[F(B,\mu,A,B)
K(\alpha-B) - F(A,\mu,A,B) K(\alpha-A)\right]\ , \label{Fp}\\
{\cal T}\circ F_A &=&\frac{1}{2\pi}F(A,\mu,A,B)K(\alpha-A)\ , \label{FA}\\
{\cal T}\circ F_B &=&-\frac{1}{2\pi}F(B,\mu,A,B)K(\alpha-B)\ ,\label{FB}
\end{eqnarray}
where ${\cal T}$ is defined in Eq.~(\ref{Tdef}).
Note that $F$ and $\stackrel{\circ}{F}$ can be written as
\begin{eqnarray}
F(\alpha,\mu,A,B) &=& {\cal T}^{-1}\circ[-{1\over{2}}\Theta(\alpha-\mu)]\ ,\\
\stackrel{\circ}{F}(\alpha,\mu,A,B) &=& {\cal T}^{-1}
\circ[-{1\over{2}} K(\alpha-\mu)]\ ,
\end{eqnarray}
where ${\cal T}^{-1}$ is the inverse operator of ${\cal T}$.
Using the linearity of Eqs.~(\ref{Fp}), (\ref{FA}), and (\ref{FB}),
$F'$, $F_A$, and $F_B$ can be obtained by acting ${\cal T}^{-1}$, which yields
\begin{eqnarray}
F'(\alpha,\mu,A,B) &=& \stackrel{\circ}{F}(\alpha,\mu,A,B)
-\frac{1}{\pi}\left[ F(B,\mu,A,B)\stackrel{\circ}{F}(\alpha,B,A,B) - \right.
\nonumber \\
&& \left. F(A,\mu,A,B)\stackrel{\circ}{F}(\alpha,A,A,B)\right]\ ,\label{F'} \\
F_A(\alpha,\mu,A,B)  &=& -\frac{1}{\pi} F(A,\mu,A,B)
\stackrel{\circ}{F}(\alpha,A,A,B) , \label{Fa}\\
F_B(\alpha,\mu,A,B)  &=& \frac{1}{\pi} F(B,\mu,A,B)
\stackrel{\circ}{F}(\alpha,B,A,B)\ . \label{Fb}
\end{eqnarray}

Using these relations in Eq.~(\ref{F1_a}) and (\ref{F2_a}), one can obtain
\begin{eqnarray}
\frac{d}{da} F(B,B,A,B) &=& \frac{2}{\pi}
F(A,B,A,B)\stackrel{\circ}{F}(B,A,A,B) \\
\frac{d}{da} F(B,A,A,B) &=& 2 \stackrel{\circ}{F}(B,A,A,B) \left[ 1 +
\frac{1}{\pi} F(A,B,A,B) \stackrel{\circ}{F}(B,A,A,B) \right]\ ,
\end{eqnarray}
from which the derivatives of $D(B)$ and $D_2(B)$ can be written as
\begin{eqnarray}
\frac{dD(B)}{da} &=& \frac{2}{\pi} \stackrel{\circ}{F}(B,A,A,B) \left\{
1+\frac{1}{\pi}\left[F(A,A,A,B)-F(A,B,A,B)\right]\right\}\ , \label{dDda}\\
\frac{dD_2(B)}{da} &=& 2\stackrel{\circ}{F}(B,A,A,B) \left\{
1+\frac{1}{\pi}\left[F(A,A,A,B)+F(A,B,A,B)\right]\right\}\ \label{dD2da} .
\end{eqnarray}
As explained in Sec.~\ref{sec3}, $F(A,A,A,B)=-F(B,B,A,B)$ and
$F(B,A,A,B)=-F(A,B,A,B)$ because the function $\Theta$ is odd and $A=-B^*$.
Using these properties in Eqs.~(\ref{dDda}) and (\ref{dD2da}), one can find
that
\begin{eqnarray}
\frac{d}{da}D(B) &=& \frac{2}{\pi} \stackrel{\circ}{F}(B,A,A,B)D(B) \\
\frac{d}{da}\left(1-\frac{D_2(B)}{\pi}\right) &=& -\frac{2}{\pi}
\stackrel{\circ}{F}(B,A,A,B) \left(1-\frac{1}{\pi}D_2(B)\right) \ .
\end{eqnarray}
Then the total derivative of $J$ is obtained straightforwardly.
\begin{eqnarray}\label{dJda}
\frac{dJ}{da} &=& \frac{dD(B)}{da}\left(1-\frac{D_2(B)}{\pi}\right) +
D(B) \frac{d}{da}\left(1-\frac{D_2(B)}{\pi}\right) \nonumber\\
&=& 0\ .
\end{eqnarray}
Equation.~(\ref{dJda}) together with $J=1$ at $a=0$ implies that
$J=1$ for any value of $a$.

\section{Modular transformation properties of $\widetilde{\cal
Z}$}\label{app3}
In this Appendix, we examine the modular covariance of the TPF for the
ASSV model in Eq.~(\ref{mpf_tw}).
It is convenient to work with the
Gaussian model TPF with the shift boundary conditions~\cite{FraSZ_Pasquer}:
\begin{equation}
{\cal Z}_{m,n}(\tau) = {\cal Z}_0 \exp\left[-\frac{\pi
g}{\tau''}|n-\tau m|^2\right]\ ,
\end{equation} where \begin{equation} {\cal Z}_0(\tau) =
\frac{g}{\tau''} \frac{1}{\eta({\sf q})\eta(\bar{\sf q})}
\end{equation} is the Gaussian model TPF with the periodic boundary
conditions. They have modular transformation properties as
\begin{equation} {\cal Z}_{m,n}(\tau) = {\cal Z}_{m,n+m}(\tau+1) =
{\cal Z}_{-n,m}(\text{\tiny $-\frac{1}{\tau}$}) \end{equation} or combining
these
\begin{equation} {\cal Z}_{m,n}(\tau) = {\cal Z}_{dm+cn,an+bm}
\left(\frac{a\tau+b}{c\tau+d}\right)
\end{equation}
where $a,b,c$, and $d$ are integers with $ad-bc=1$.

The generalized Coulombic partition function defined in Eq.~(\ref{mpf_p})
is written in terms of $Z_{m,n}$ by applying the Poisson sum formula:
\begin{equation}
\sum_{m=-\infty}^\infty f(m-l) = \sum_{m=-\infty}^\infty e^{2\pi i
ml}g(2\pi m) \end{equation} with \begin{equation} g(k) \equiv
\int_{-\infty}^\infty dx\, f(x)e^{ikx}\ . \end{equation}
First, Eq.~(\ref{mpf_p}) can be written as
\begin{eqnarray}
\widetilde{\cal Z}_{\text{gC}}(\alpha,\beta;l,l'|\tau) &=&
\frac{1}{|\eta({\sf q})|^2}  \sum_{n\in {\bf Z}}
\exp\left[-\pi g\tau''(n-\beta)^2-2\pi il'(n-\beta)-2\pi i\alpha l \right]
\nonumber \\
&& \times \sum_{m\in{\bf Z} } \exp\left[-\frac{\pi\tau''}{g}(m-l)^2+2\pi
i(\tau'(n-\beta)-\alpha)(m-l)\right]\ . \label{Ze}
\end{eqnarray}
Then, applying the Poisson sum formula to the sum over $m$ in
Eq.~(\ref{Ze}), one can express $\widetilde{\cal Z}_{\text gC}$ in terms of
${\cal Z}_{n,m}$ as
\begin{equation} \widetilde{\cal Z}_{\text gC}(\alpha,\beta;l,l'|\tau) =
{\cal Z}_0(\tau)
\sum_{m,n\in{\bf Z}} e^{-2\pi i(m+\alpha)l-2\pi il'(n-\beta)} {\cal
Z}_{n-\beta,m+\alpha}(\tau)\ . \end{equation}
And the TPF in Eq.~(\ref{mpf_tw}) becomes
\begin{eqnarray}
\widetilde{\cal Z}(\rho_L,\rho_D;l,l';M,N|\tau) &=&
\exp\left[\pi iMl(2\rho_L-1)-\pi iNl' (2\rho_D-1)\right] {\cal Z}_0(\tau)
\nonumber \\
&& \times \sum_{m,n\in{\bf Z}} e^{-2\pi i(m+\alpha)l-2\pi
il'(n-\beta)} {\cal Z}_{n-\beta,m+\alpha}(\tau) \end{eqnarray}
where $\alpha\equiv\{M\rho_L\}$ and $\beta\equiv\{N\rho_D\}$ where $\{x\}$
denotes the fractional part of $x$.

Under the transformation $\tau\rightarrow\tau+1$,
\begin{eqnarray}
\widetilde{\cal Z}_{\text{gC}}(\alpha,\beta;l,l'|\tau) &=& {\cal Z}_0(\tau)
\sum_{m,n\in{\bf Z}}
e^{-2\pi il(m+\alpha)-2\pi il'(n-\beta)} {\cal
Z}_{n-\beta,m+n+\alpha-\beta}(\tau+1) \nonumber \\
&=&  {\cal Z}_0(\tau+1) \sum_{m,n\in{\bf Z}}
e^{-2\pi il(m+\alpha-\beta)-2\pi i(l'-l)(n-\beta)} {\cal
Z}_{n-\beta,m+\alpha-\beta}(\tau+1) \nonumber \\
&=& \widetilde{\cal Z}_{\text{gC}}(\alpha-\beta,\beta;l,l'-l|\tau+1)\ ,
\end{eqnarray}
which means that
\begin{equation}
\widetilde{\cal Z}_{\text{gC}}(\alpha,\beta;l,l'|\tau+1) =
\widetilde{\cal Z}_{\text{gC}}(\alpha+\beta,\beta;l,l'+l|\tau) \ .
\end{equation}
So, the TPF $\widetilde{\cal Z}$ transforms under the transformation
$\tau\rightarrow\tau +1$ as
\begin{equation}
\widetilde{\cal Z}(\rho_L,\rho_D;l,l';M,N|\tau+1)
= \widetilde{\cal Z}(\rho_L',\rho_D;l,l'+l;M+N,N|\tau) \end{equation}
where
\begin{equation} \rho_L' = \frac{M\rho_L+N\rho_D}{M+N} .\end{equation}

Similarly, $\widetilde{\cal Z}_{\text{gC}}$ transforms under the
transformation $\tau\rightarrow -1/\tau$ as
\begin{eqnarray}
\widetilde{\cal Z}_{\text{gC}}(\alpha,\beta;l,l'|\tau) &=& {\cal Z}_0(\tau)
\sum_{m,n\in{\bf Z}} e^{-2\pi i(m+\alpha)l-2\pi il'(n-\beta)} {\cal
Z}_{-m-\alpha,n-\beta}(\text{\tiny $-\frac{1}{\tau}$}) \nonumber \\
&=&  {\cal Z}_0(\text{\tiny $-\frac{1}{\tau}$}) \sum_{m,n\in{\bf Z}}
e^{-2\pi il'(n-\beta)-2\pi i(-l)(m-\alpha)} {\cal Z}_{m-\alpha,n-\beta}
(\text{\tiny $-\frac{1}{\tau}$}) \nonumber \\
&=& \widetilde{\cal Z}_{\text{gC}}(-\beta,\alpha;l',-l|
\text{\tiny $-\frac{1}{\tau}$})\ ,
\end{eqnarray}
which means that
\begin{equation}
\widetilde{\cal Z}_{\text{gC}}(\alpha,\beta;l,l'|
\text{\tiny $-\frac{1}{\tau}$}) =
\widetilde{\cal Z}_{\text{gC}}(\beta,-\alpha;-l',l|\tau) \ .
\end{equation}
{}From this, one can also find that
\begin{equation}
\widetilde{\cal Z}(\rho_L,\rho_D;l,l';M,N| \text{\tiny $-\frac{1}{\tau}$})
= \widetilde{\cal Z}(\rho_L',\rho_D';-l',l;N,M|\tau)
\end{equation}
where
\begin{equation} \rho_L' = \rho_D\ ,\ \ \ \rho_D' = 1-\rho_L .\end{equation}

\end{appendix}

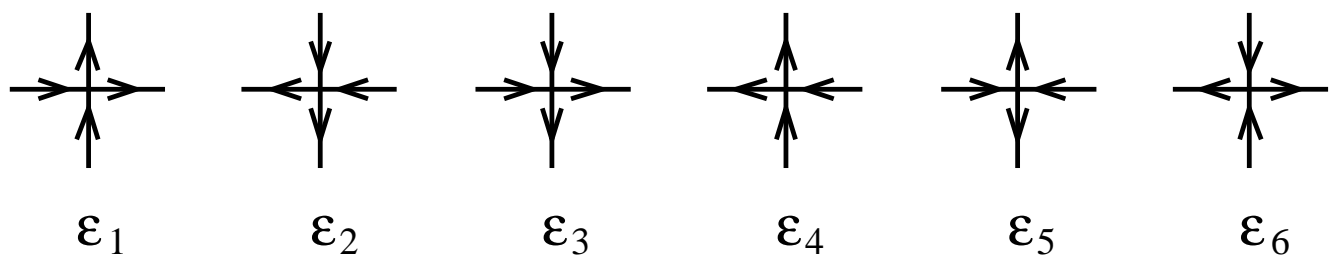
\begin{figure}
\caption{Six vertex configurations.}
\end{figure}

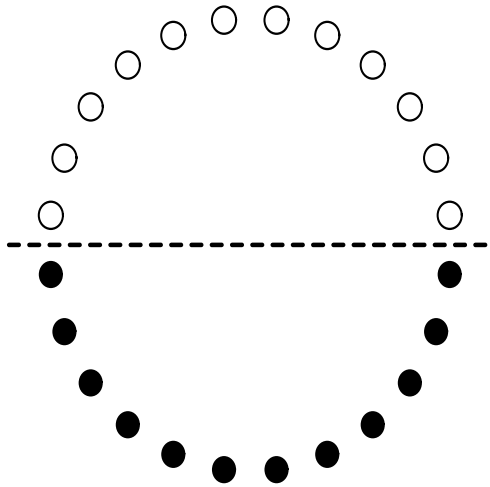
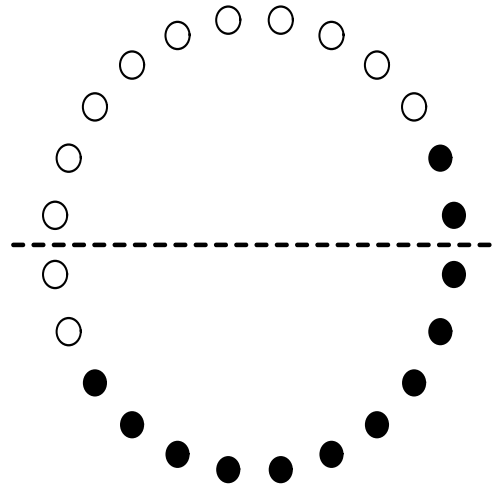
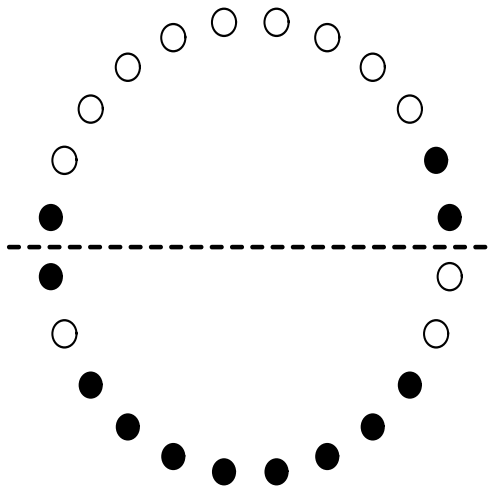
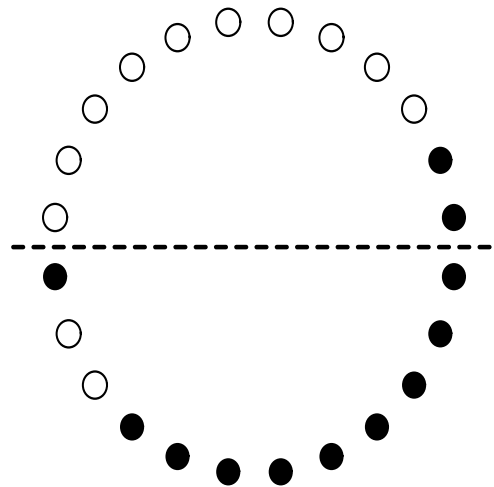
\begin{figure}
\caption{A few examples of particle-hole configuration for $N=24$ and
$Q=12$. We denote the values of $I_j$ by the angular coordinate of
equi-spaced small circles on a unit circle.
Closed circles denote the occupied positions. (a) depicts the
the ground state, (b) the $m=2$-shifted state,
while (c) and (d) shows excited states from (a) and (b), respectively.}
\end{figure}

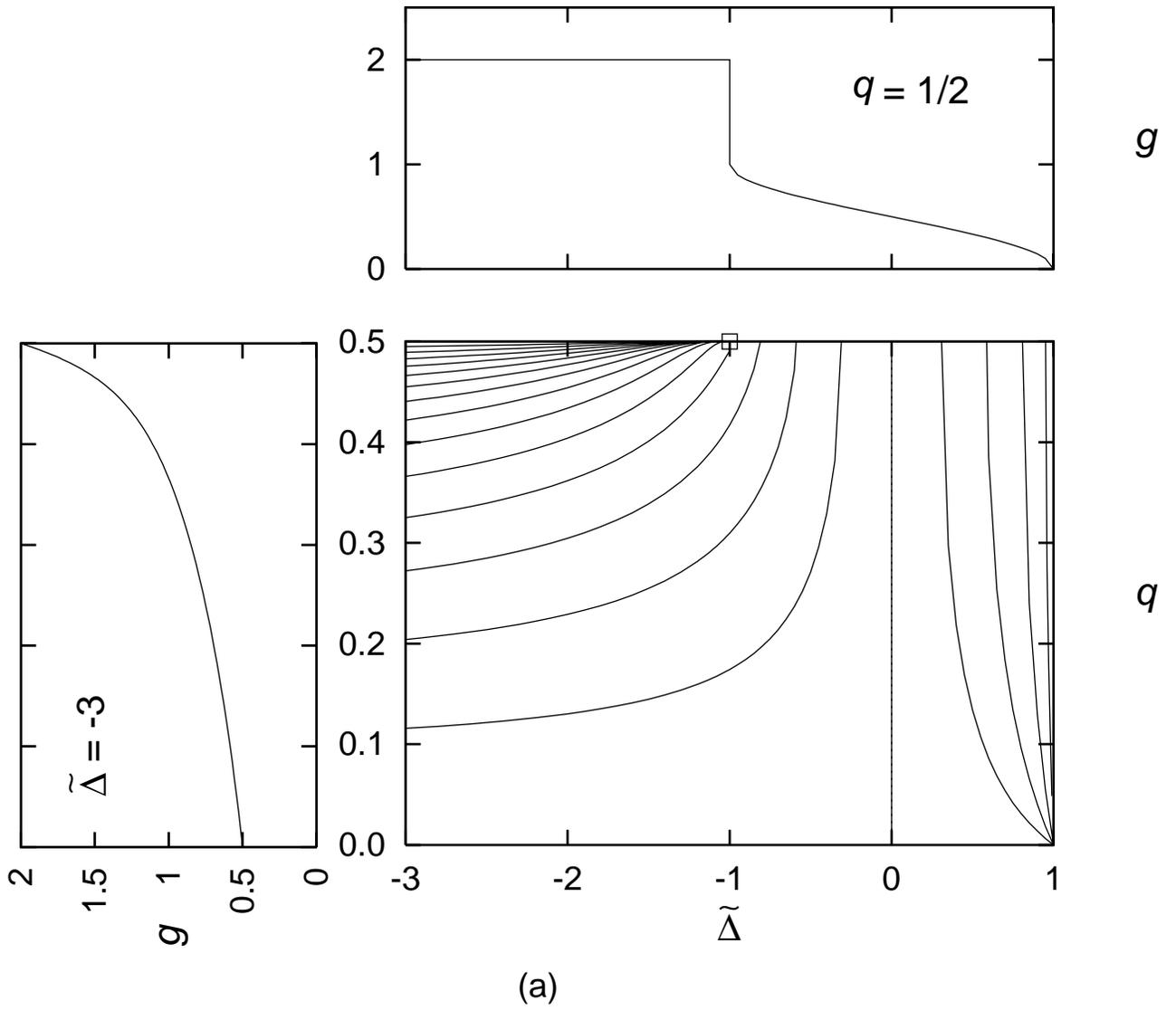
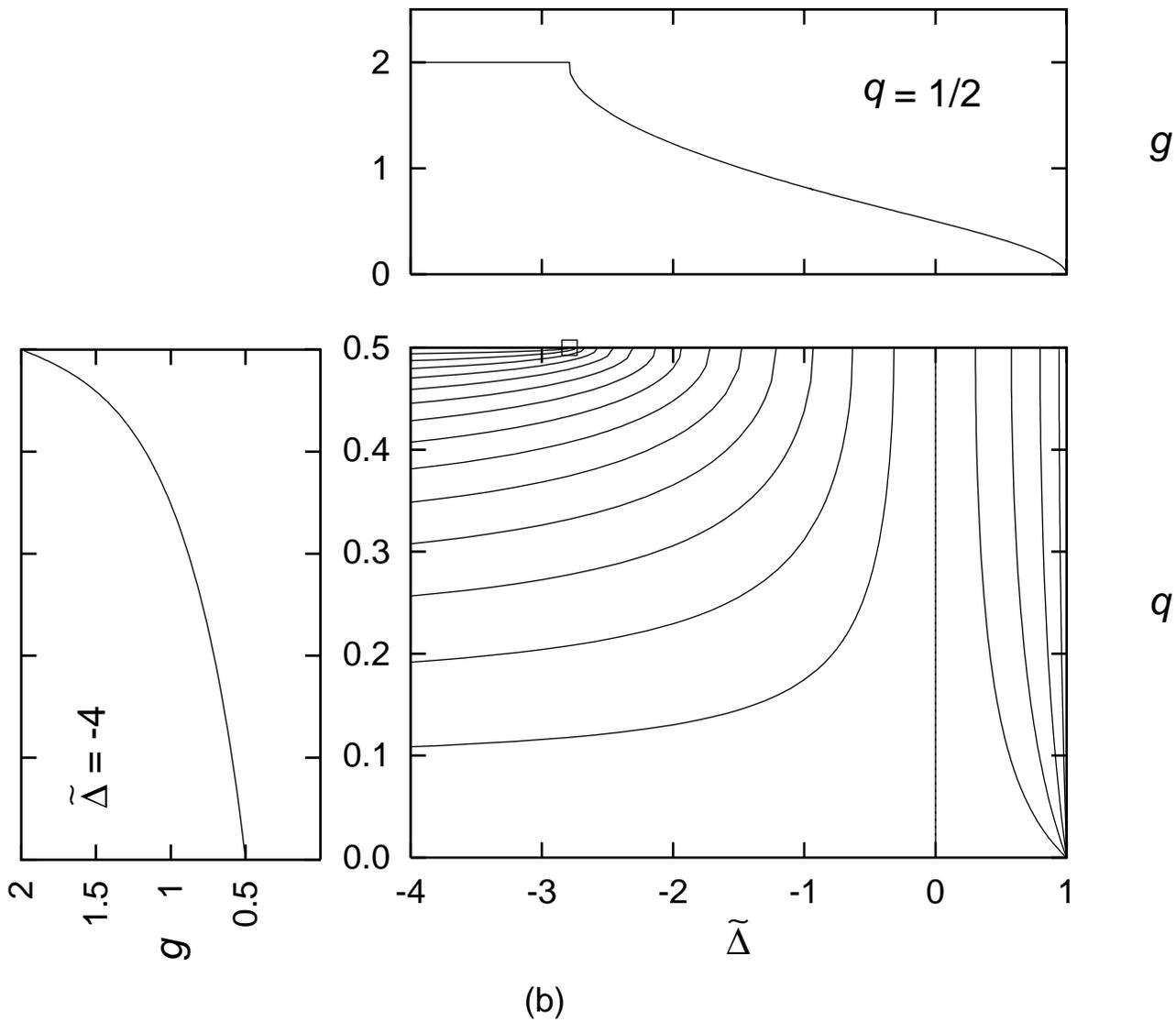
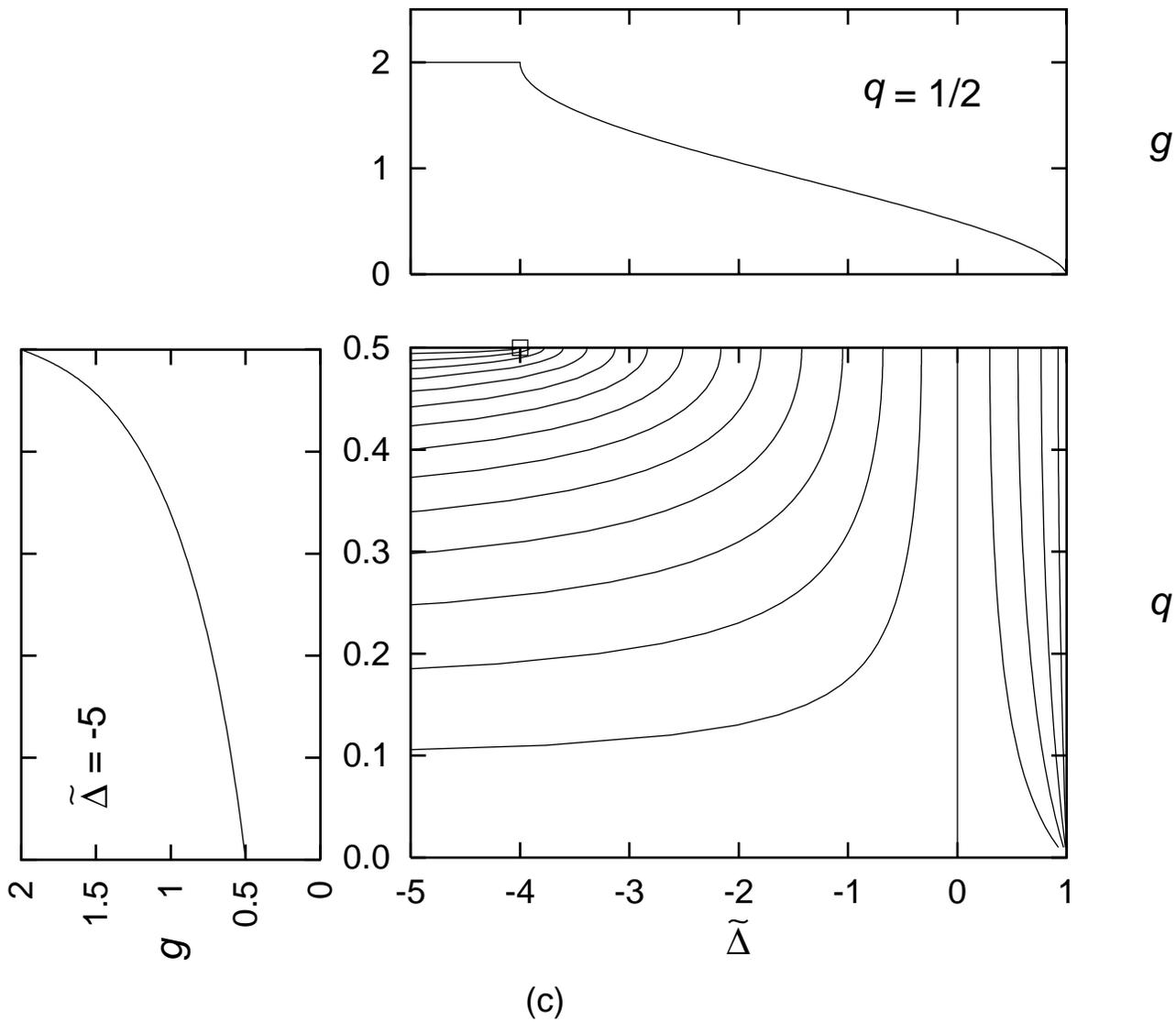
\begin{figure}
\caption{The constant-$g$ lines are plotted in the $(\widetilde{\Delta},q)$
plane for $\tanh(2h)=0$ in (a), ${0.5}$ in (b), and ${1}$ in (c).
These are obtained from the
numerical solutions of Eq.~(\protect{\ref{Dint}}).
On $\widetilde{\Delta}=0$ and $q=0$ lines, $g=1/2$.
On $\widetilde{\Delta}=1$ line $g=0$.
The value of $g$ increases in steps of $0.1$ from right to left and reaches
$2$ at $q=1/2$ and $\widetilde{\Delta}<\widetilde{\Delta}_c$. The figures
at the top show $g$ as a function of $\widetilde{\Delta}$ at
$q=1/2$ and the figures at left show $g$ as a function of $q$
at $\widetilde{\Delta}=-3$ in (a), ${-4}$ in (b), and ${-5}$ in (c).
The rectangular symbols denote the phase-transition points
$\widetilde{\Delta}_c$ from the critical phase to the $q=1/2$ ordered phase.}
\end{figure}

\end{document}